\documentstyle[12pt]{article}
\begin{document}
\parskip=5pt plus 1pt minus 1pt

\begin{flushright}
{\bf KEK Preprint 97-22}\\
{\bf DPNU-97-26}
\end{flushright}
\begin{flushright}
{May 1997}
\end{flushright}

\vspace{0.2cm}
\begin{center}
{\Large\bf Quark Mass Matrices in Superstring Models}
\end{center}
\vspace{0.5cm}

\begin{center}
{\bf Tatsuo Kobayashi} \footnote{ Electronic 
address: Kobayast@tanashi.kek.jp} \\
{\it Institute of Particle and Nuclear Studies, \\
High Energy Accelerator Research Organization, Tanashi, Tokyo 188, Japan}
\end{center}

\begin{center}
{\bf Zhi-zhong Xing} \footnote{ Electronic 
address: Xing@eken.phys.nagoya-u.ac.jp} \\
{\it Department of Physics, Nagoya University, 
Chikusa-ku, Nagoya 464-01, Japan}
\end{center}
\vspace{1.cm}

\begin{abstract}
Four simple but realistic patterns of quark mass matrices
are derived from orbifold models of superstring theory
in the absence of gauge symmetries.
Two of them correspond to the Ramond-Roberts-Ross types,
which have five texture zeros in up and down quark sectors.
The other two, with four texture zeros, preserve the structural parallelism 
between up and down sectors. The phenomenological consequences of these mass 
matrices 
on flavor mixings and $CP$ violation are analyzed at the weak scale. With
the same input values of quark mass ratios, we find that only one or two of the 
four patterns can be in good agreement with current experimental data.
\end{abstract}

\vspace{1cm}
\begin{center}
({\sl Accepted for publication in Int. J. Mod. Phys. A})
\end{center}

\newpage

\section{Introduction}

The origin of fermion masses is one of the most important 
problems in particle physics. To date
much work has been devoted to understanding this puzzle. 
One of the plausible mechanisms for fermion mass generation is to use 
nonrenormalizable couplings such as $HQ_iq_j(\theta /M)^{n_{ij}}$.
When suitable fields like $\theta$ develop their vacuum 
expectation values (VEVs), these nonrenormalizable couplings could 
provide effective Yukawa couplings 
$HQ_iq_j(\langle \theta \rangle /M)^{n_{ij}}$ with a hierarchical 
structure \cite{qmass}.
A realistic hierarchy can be derived from certain types of structures of 
the power parameters $n_{ij}$.

The structure of these powers is determined by underlying theories
with underlying symmetries.
For example, gauge symmetries as well as global symmetries can be used 
to derive fermion mass matrices \cite{qmass,IR,qmass2}.

Superstring theory is a promising candidate for the unified theory 
of all interactions including gravity.
Superstring theory has proper selection rules for nonrenormalizable 
couplings in addition to gauge symmetries of field theory.
These proper symmetries in superstring theory, stringy symmetries,  are 
originated from 
structure of a six-dimensional compactified space of string vacua.
They can lead to realistic fermion mass matrices, which may 
differ from those derived from gauge symmetries
\footnote{In the approach of gauge symmetries 
sfermion masses gain D-term contributions to soft scalar masses through 
supersymmetry breaking \cite{Dterm}. Such D-term contributions
should be treated carefully so that they  
do not break degeneracy between squark masses, as required by present
experiments of flavor changing neutral currents \cite{fcnc}.}.
Actually some efforts have been made to derive fermion mass matrices 
by use of stringy symmetries in Refs. \cite{stringm}--\cite{stringm22}.
Refs. \cite{stringm,stringmcy} discuss fermionic construction 
of 4D string models and Calabi-Yau models, where stringy symmetries are 
discrete symmetries. In comparison, Refs. \cite{stringm21,stringm22} 
discuss orbifold models, which have complicated selection rules \cite{nr,NR}.
A systematic study of possible patterns of quark mass matrices that can
be derived from stringy symmetries, however, has been lacking in the
literature.

In this work we investigate various possibilities to obtain simple but realistic  
quark mass matrices from orbifold models by nontrivially extending the analysis 
in 
Refs. \cite{stringm21,stringm22}, where only one type of symmetric quark 
mass matrices was discussed. Following the principle of simplicity,
here we concentrate only on symmetric quark mass matrices with five or 
four texture zeros. It has been shown by Ramond, Roberts
and Ross (RRR) that symmetric quark mass matrices with more than five texture 
zeros are not realistic \cite{RRR}, i.e., they cannot fit current experimental
data on quark mixings and $CP$ violation. Note that symmetric quark mass 
matrices 
with five texture zeros can be classified into five types, the so-called RRR 
patterns
\cite{RRR}. We show that only two of the five RRR patterns can be naturally 
derived from orbifold models. Since mass matrices with four texture zeros 
include 
more degrees of freedom, we further impose the up-down structural parallelism 
on them to obtain relatively simple patterns. Then we can find that there exist
only two types of realistic quark mass matrices with four texture zeros and 
up-down parallel structures. The phenomenological consequences of the obtained 
quark mass matrices are also analyzed, from the string scale to the weak scale, 
by use of renormalization-group equations in the framework of the minimal 
supersymmetric standard model (MSSM).

The remaining parts of this paper are organized as follows.
In section 2 we first give a brief review of selection rules of orbifold models.
Using the selection rules of couplings we derive
some typical patterns of quark mass matrices, and then calculate their
mass eigenvalues and flavor mixing matrices at the string scale.
In section 3 the obtained patterns of mass matrices are confronted with 
current experimental data at the weak scale. Section 4 is devoted to a summary
with some concluding remarks.

\section{Quark mass matrices in orbifold models}
\setcounter{equation}{0}

\begin{center}
{\large\bf A. ~ Orbifold models}
\end{center}

The orbifold construction is one of the simplest and most interesting
constructions to derive four-dimensional string vacua \cite{Orbi}.
In orbifold models, string states consist of the bosonic string on
the four-dimensional space-time and a six-dimensional orbifold,
as well as their right-moving superpartners and left-moving gauge parts.
The right-moving fermionic parts are bosonized, and momenta of
bosonized fields span an SO(10) lattice.
An orbifold is obtained through the division of a six-dimensional space
$R^6$ by a six-dimensional lattice and its automorphism $\theta$.
Closed strings on the orbifold are classified into untwisted and
twisted types.
For the $\theta^k$-twisted sector $T_k$, the string coordinate
on the orbifold, $x_\nu$ ($\nu=1 \sim 6$), satisfies the following 
boundary condition:
\begin{eqnarray}
 x_\nu(\sigma=2 \pi) \; = \; (\theta^kx)_\nu(\sigma=0) ~ + ~ e_\nu \; ,
\label{bc}
\end{eqnarray}
where $e_\nu$ is a lattice vector.
A zero-mode of this string satisfies the same condition as Eq. (\ref{bc}) 
and it is called a fixed point.
The fixed point is represented in terms of its space group 
element $(\theta^k,e_\nu)$; and we have the corresponding 
six-dimensional ground state $|(\theta^k,e_\nu) \rangle$.
All fixed points in $T_k$ are not fixed under $\theta$.
To obtain $\theta$-eigenstates, we have to take linear combinations
of states corresponding to fixed points of $\theta^k$ as \cite{KO1,KO2} 
\begin{eqnarray}
|(\theta^k,e_\nu) \rangle + \gamma^{-1}|\theta(\theta^k,e_\nu) \rangle 
+ \cdots + \gamma^{-(m-1)}|\theta^{m-1}(\theta^k,e_\nu) \rangle \; .
\label{state}
\end{eqnarray}
Here $\theta^m$ denotes the smallest twist fixing $(\theta^k,e_\nu)$ itself. 
Thus we have $m<k$.
These linear combinations have eigenvalues
$\gamma =\exp[{\rm i}2 \pi n/m]$ under the $\theta$-twist with an integer $n$.
We denote the $T_k$ sector with the $\theta$-eigenvalue $\gamma$ as 
$T_{k(\gamma)}$, when we specify the $\theta$-eigenvalue.
We take a complex basis $(X_i,\overline X_i)$ $(i=1 \sim 3)$ for
the compactified space, e.g., $X_1=x_1+ {\rm i} x_2$.
Oscillated states in $T_k$ are created by $\partial X_{i(k)}$ and
$\partial \overline X_{i(k)}$ on the ground states given in (\ref{state}).
The twisted sectors have shifted SO(10) momenta.
Every shifted SO(10) momentum of massless $T_k$ states has been
shown explicitly in Refs. \cite{KO1,KO2}.

Couplings are calculated by using vertex operators $V_a$
corresponding to states \cite{FMS,Yukawa}.
Vertex operators consist of several parts, i.e., the four-dimensional 
part, the six-dimensional ground state of the $T_k$ sector as 
(\ref{state}), oscillators on it, the bosonized SO(10) part,  
the gauge part, and the ghost part.
Nonvanishing couplings are invariant under a symmetry of each part.
Coupling terms are allowed if they are gauge invariant and space-group 
invariant. 
The latter implies that a product of space-group elements $(\theta^k_a,e_a)$ for 
coupling states should satisfy $\prod_a (\theta^k_a,e_a)=(1,0)$ up to 
the conjugacy class 
\footnote[2]{For the selection rule due to the space group, see Ref. \cite{KO2}
for some details.}.
Furthermore, a product of eigenvalues $\gamma_a$ should satisfy 
$\prod_a \gamma_a=1$.
In addition, the SO(10) momentum and the ghost number should be conserved. 
The corresponding correlation function $\langle V_1 \cdots V_n \rangle$
should be invariant under a $Z_N$ rotation of oscillators as
\begin{eqnarray}
\partial X_{i(k)} \; \rightarrow \; e^{{\rm i} 2 \pi kv^i} \partial X_{i(k)} \; 
,
\end{eqnarray}
where $e^{{\rm i} 2 \pi v^i}$ are eigenvalues of $\theta$ in the
complex basis $X_i$. Note that even vertex operators corresponding to 
non-oscillated massless states include oscillators, when we change their 
pictures from  
the $-1$ or $-1/2$ picture \cite{FMS}.
These selection rules have been discussed in Ref. \cite{NR}.

$Z_{6}$-II orbifold models have most twisted sectors among 
$Z_N$ orbifold models, whose six-dimensional orbifolds are constructed 
as products of two-dimensional orbifolds.
Thus one can expect that $Z_{6}$-II orbifold models are simpler but possess
various types of couplings, which could be useful to 
obtain realistic mass matrices. For this reason,
we concentrate our attention upon $Z_{6}$-II orbifold models in the following.

The $Z_6$-II orbifold has eigenvalues $v_i=(2,1,-3)/6$.
Massless matter states in $Z_{6}$-II orbifold models correspond to 
$T_1$, $T_2$, $T_3$ and $T_4$ sectors \cite{KO1,KO2,Z6}.
The other twisted sectors correspond only to antimatter states.
Allowed nonrenormalizable couplings in $Z_6$-II orbifold models are listed
in Table 1, where PGI means the point group invariance.
\begin{table}
\caption{Allowed nonrenormalizable couplings in $Z_6$-II orbifold 
models.}
\vspace{-0.2cm}
\begin{center}
\begin{tabular}{cc} \\ \hline\hline \\
~~~~~~ Coupling ~~~~~~ & ~~~~~~~~~~ Condition ~~~~~~~~~~ \\ \\ \hline \\
  $T_2^{3\ell}T_3^{2m}$         & $\ell >0 \; , \;\; m >0$ \\ \\
  $T_4^{3\ell}T_3^{2m}$         & $\ell >0 \; , \;\; m >0$  \\ \\
  $T_1^{2\ell}T_2^mT_4^n$       & $\ell =2p+1 \; , \;\; 2\ell +2m+n=3q \; , \;\;
                                    \ell >0 \; , \;\; m>0 \; , \;\; n>0 \; , \; 
\; {\rm PGI}$ \\ \\
   $T_1^\ell T_2^mT_3^nT_4^p$   & $\ell >0 \; , \;\; n>0 \; , \;\;
                                    m \; {\rm or } \; p>0 \; , \;\; {\rm PGI}$ 
\\ \\
\hline\hline
\end{tabular}
\end{center}
\end{table}

\begin{center}
{\large\bf B. ~ Mass matrices with up-down parallelism}
\end{center}

In general, the underlying theory like supergravity or superstring theory 
has nonrenormalizable couplings as  
\begin{equation}
h_{{\rm u}ij}H_2Q_iu_j(\theta_{\rm u}/M_2)^{n_{ij}} \; , ~~~~~~~~
h_{{\rm d}ij}H_1Q_id_j(\theta_{\rm d}/M_1)^{n'_{ij}} \; ,
\end{equation}
where $Q_i$ is the SU(2) doublet of quark fields, 
$u_j$ ($d_j$) denotes the up-type (down-type) SU(2) singlet of 
quark fields, and $H_{2,1}$ are 
the Higgs fields for the up and down sectors.
Here $h_{{\rm u}ij}$ and $h_{{\rm d}ij}$ denote coupling strengths, 
which can be calculated within the framework of superstring theory.
Their magnitudes are of $O(1)$ in most cases.
When the fields $\theta_{\rm u,d}$ develop VEVs, these couplings become 
Yukawa couplings with suppression factors 
$\varepsilon_{\rm u}=(\langle\theta_{\rm u}\rangle/M_{2})^{n_{ij}}$ and 
$\varepsilon_{\rm d}=(\langle\theta_{\rm d}\rangle/M_{1})^{n'_{ij}}$. 
They can lead to a hierarchical structure in the fermion mass matrices.
In general, we expect $\varepsilon_{\rm u} \neq \varepsilon_{\rm d}$. 
For example, the mixing between light 
and heavy Higgs fields leads to 
$\varepsilon_{\rm u} \neq \varepsilon_{\rm d}$ \cite{IR,RRR}.

Using allowed nonrenormalizable couplings in $Z_6$-II orbifold models, 
we can obtain simple and realistic quark mass matrices.
We first study the possibility for symmetric quark mass matrices 
with up-down structural parallelism and four texture zeros.
Here the texture zero does not mean that the matrix element is completely 
vanishing; instead it means that the matrix element is remarkably
suppressed or vanishingly small. 

We assign Higgs fields $H_1$ and $H_2$ to the $T_{4(\gamma =1)}$ 
sector. The first, second and third families of both up and down 
quarks are assigned to $T_{2(\gamma =1)}$, $T_{3(\gamma =1)}$ and $T_1$, 
respectively.
In this assignment we assume fields with $T_1$, $T_{2(\gamma =-1)}$ and 
$T_{4(\gamma =-1)}$ sectors to develop VEVs. Then we obtain the following quark 
mass 
matrices \cite{stringm21,stringm22}:
\begin{eqnarray}
{\rm Pattern ~ S1:} ~~~~~~~~~~~~~~ 
M_{\rm u,d}=c_{\rm u,d}
\pmatrix{
0 & \varepsilon_{\rm u,d}^3 & 0 \cr
\varepsilon_{\rm u,d}^3 & \varepsilon_{\rm u,d}^2 
& \varepsilon_{\rm u,d}^2 \cr
0 & \varepsilon_{\rm u,d}^2 & 1 \cr
} ~~~~~~
\label{lr1}
\end{eqnarray}
with $h_{{\rm u}ij}/h_{{\rm u}33}$ ($h_{{\rm d}ij}/h_{{\rm d}33}$) 
of $O(1)$. Here $c_{\rm u}=h_{{\rm u}33}\langle H_2 \rangle$ and 
$c_{\rm d}=h_{{\rm d}33}\langle H_1 \rangle$.
The (1,2), (2,2) and (2,3) elements are originated from 
$(T_2T_3T_4)T_1^3$, $(T_3^2T_4)T_{4(\gamma =-1)}^2$ and 
$(T_1T_3T_4)T_{2(\gamma =-1)}^2$ couplings, respectively.
The (1,1) or (1,3) element does not completely vanish, but it is
sufficiently suppressed in comparison with its neighboring elements.
For example, the (1,3) element can be obtained as $\varepsilon_{\rm u,d}^9$ by 
$(T_1T_2T_4)T_1^9$, if we assign fixed points in a certain way.

We are able to obtain another type of quark mass matrices with the 
up-down parallel structure.
We assign both of $H_1$ and $H_2$ to $T_{4(\gamma =1)}$, and assign 
the first, second and third families of up and down quarks to 
$T_{3(\gamma=1)}$, $T_{3(\gamma=\omega)}$ and $T_1$, respectively, 
where $\omega=e^{{\rm i} 2\pi /3}$.
In this assignment, $T_1$, $T_{3(\gamma=\omega)}$ and 
$T_{4 (\gamma =1)}$ fields are assumed to develop VEVs.
Then we get quark mass matrices of the type
\begin{eqnarray}
{\rm Pattern ~ S2:} ~~~~~~~~~~~~~~
M_{\rm u,d}=c_{\rm u,d}
\pmatrix{
0 & \varepsilon_{\rm u,d}^4 & 0 \cr
\varepsilon_{\rm u,d}^4 & \varepsilon_{\rm u,d}^3 
& \varepsilon_{\rm u,d}^3 \cr
0 & \varepsilon_{\rm u,d}^3 & 1 \cr
}. ~~~~~~
\label{ud2}
\end{eqnarray}
Here the (2,2) and (2,3) elements are originated from 
$T_1T_{3(\gamma =\omega)}^3T_{4(\gamma =1)}^2$ coupling, while 
the (1,2) element is due to 
$T_{3(\gamma =1)}T_{3(\gamma=\omega)}^3T_{4(\gamma =1)}^3$ coupling.

Note that those quark mass matrices with four texture zeros 
in entries different from (1,1) and (1,3) are phenomenologically unrealistic.
Thus patterns S1 and S2 should be the only candidates of realistic and simple 
quark mass matrices with four texture zeros and up-down structural parallelism.
Giving up the up-down structural parallelism, one can indeed derive many
patterns of quark mass matrices with four texture zeros. Such a work is
less interesting in both theory and phenomenology, because there may
exist much simpler patterns with five texture zeros \cite{RRR} 
(as one can see later on).

\begin{center}
{\large\bf C. ~ Mass matrices of Ramond-Roberts-Ross types}
\end{center}

Five types of symmetric quark mass matrices with 
five texture zeros, the so-called RRR patterns, have been obtained in Ref. 
\cite{RRR}.
The down mass matrices in three RRR patterns take the form like 
(2.6). Thus one can in principle derive these three RRR patterns from orbiford 
models 
by choosing the appropriate assignment for the up-type quarks.

For example, we can assign $u$, $c$ and $t$  
quarks to $T_{3(\gamma=\omega)}$, $T_{3(\gamma=\omega^2)}$ and 
$T_1$, respectively, without changing the assignment of other fields.
Then we have 
\begin{equation}
{\rm Pattern ~ A1:} ~~~~~~~~~~
M_{\rm u} \; = \; c_{\rm u}
\pmatrix{
0 & \varepsilon_{\rm u}^3 & 0 \cr
\varepsilon_{\rm u}^3 & \varepsilon_{\rm u}^2 
& 0 \cr
0 &0 & 1 \cr
}, \quad 
M_{\rm d} \; = \; c_{\rm d}
\pmatrix{
0 & \varepsilon_{\rm d}^4 & 0 \cr
\varepsilon_{\rm d}^4 & \varepsilon_{\rm d}^3 
& \varepsilon_{\rm d}^3  \cr
0 &\varepsilon_{\rm d}^3   & 1 \cr
}.
\label{rrr1}
\end{equation}
This pattern corresponds to the first RRR-type mass matrices \cite{RRR}.
The (2,2) element in $M_{\rm u}$ is originated from 
$T_{3(\gamma =\omega)}T_{3(\gamma =\omega^2)}T_4T_1^2$ and/or 
$T_{3(\gamma =\omega)}T_{3(\gamma =\omega^2)}T_4^3$; while 
the (1,2) and (2,1) elements are due to 
$T_{3(\gamma =\omega)}^3T_4^2T_1$ and 
$T_3T_{3(\gamma =\omega)}T_{3(\gamma =\omega^2)}T_4^2T_1$, respectively.
The other elements of $M_{\rm u}$, e.g., the (2,3) and (3,2) elements, 
can be sufficiently suppressed if fixed points are chosen in a proper way.

One can also obtain the third RRR-type mass matrices in a similar way.
For example, $u$, $c$ and $t$ quarks may be assigned to $T_{3(\gamma=1)}$, 
$T_{3(\gamma=\omega^2)}$ and $T_1$, respectively, without 
changing the assignment of other fields.
The resultant quark mass
matrices take the form
\begin{equation}
{\rm Pattern ~ A2:} ~~~~~~~~~~
M_{\rm u} \; = \; c_{\rm u}
\pmatrix{
0 & 0& \varepsilon_{\rm u}^3  \cr
0 & \varepsilon_{\rm u}^3 & 0 \cr
\varepsilon_{\rm u}^3 &0 & 1 \cr
}, \quad 
M_{\rm d} \; = \; c_{\rm d}
\pmatrix{
0 & \varepsilon_{\rm d}^4 & 0 \cr
\varepsilon_{\rm d}^4 & \varepsilon_{\rm d}^3 
& \varepsilon_{\rm d}^3  \cr
0 &\varepsilon_{\rm d}^3   & 1 \cr
}.
\end{equation}
Here the (2,2) and (1,3) elements are originated from 
$T_{3(\gamma =\omega^2)}T_{3(\gamma =\omega)}T_4T_1^2$ and 
$T_1^2T_{3(\gamma =1)}^2T_4$ couplings, respectively.
The other matrix elements in $M_{\rm u}$ may be remarkably suppressed through
a proper choice of fixed points.

It is difficult to obtain the
other three RRR patterns of mass matrices from $Z_6-$II orbifold models.
One obvious reason is that those three patterns involve more complicated 
hierarchies.
For example, the up mass matrix in the fourth RRR pattern reads \cite{RRR}:
\begin{equation}
M_{\rm u} \; \sim \; c_{\rm u}
\pmatrix{
0 & \lambda^6 & 0  \cr
 \lambda^6  &  \lambda^4 &  \lambda^2  \cr
0 & \lambda^2  & 1 \cr
},
\label{RRR4}
\end{equation}
where $\lambda =0.22$ is the Cabibbo angle. Clearly $M_{\rm u}$ in (2.9)
contains five hierarchies: $1$, $\lambda^2$, $\lambda^4$, $\lambda^6$ 
and $0$. Such a complicated hierarchical structure cannot be easily derived from 
orbifold models by only using stringy selection rules, 
because the number of twisted sectors is limited.
In the above-mentioned mechanism of $Z_6-$II orbifold models, 
it is also difficult to make the (2,2) element of a mass matrix 
vanishingly small in comparison with its nonvanishing (2,3) and (3,2) elements.
For example, there is not a straightforward way to reproduce the second RRR 
pattern 
which includes the following up mass matrix \cite{RRR}: 
\begin{equation}
M_{\rm u} \; \sim \; c_{\rm u}
\pmatrix{
0 & \lambda^6 & 0  \cr
 \lambda^6  &  0 &  \lambda^2  \cr
0 & \lambda^2  & 1 \cr
}.
\end{equation}
Similarly, we find that the popular Fritzsch-type mass matrices 
\cite{Fritzsch78} cannot be easily derived from string models by purely using 
stringy selection rules.

\begin{center}
{\large\bf D. ~ Mass eigenvalues and flavor mixings}
\end{center}

In this subsection we first calculate mass eigenvalues for the
four types of quark mass matrices obtained above, and then
derive the corresponding flavor mixing matrices at the
string scale $M_S$. 

Up to now we have taken $h_{{\rm u}ij}/h_{{\rm u}33}=1$ and 
$h_{{\rm d}ij}/h_{{\rm d}33}=1$ .
Within the framework of superstring theory we can calculate 
magnitudes of these couplings.
The coupling strengths $h_{{\rm u}ij}$ (or $h_{{\rm d}ij}$) are obtained
as $h_{{\rm u}ij}\sim \exp(-a_{ij}T)$ 
(or $h_{{\rm d}ij}\sim \exp(-a'_{ij}T)$), where $T$ is the moduli 
parameter representing the size of six-dimensional compactified space 
and $a_{ij}$ (or $a'_{ij}$) is a constant depending on the
combination of fixed points for couplings \cite{Yukawa} 
\footnote{Similarly the $CP$-violating phases can be introduced 
into some elements in the case of nonvanishing background antisymmetric 
tensors \cite{anti}, although without such antisymmetric tensors 
$CP$ is a nice symmetry of superstring -- the relevant couplings are always real 
except trivial phases \cite{CP}.}.
The factors $\exp(-a_{ij}T)$ in the mass matrix elements are generally 
different from one another. 
These factors seem to be of $O(1)$. Note that the (1,2) and 
(2,3) elements of Pattern S1 have different powers of $\varepsilon_{\rm u,d}$ 
even in the absence of $h_{{\rm u}ij}$ (or $h_{{\rm d}ij}$),
thus the effect of $h_{{\rm u}12}/h_{{\rm u}23}$ 
(or $h_{{\rm d}12}/h_{{\rm d}23}$) is not important.
Indeed the deviation of $h_{{\rm u}12}/h_{{\rm u}23}$ 
(or $h_{{\rm d}12}/h_{{\rm d}23}$) from unity can be absorbed 
by a redefinition of the basic parameter $\varepsilon_{\rm u}$
(or $\varepsilon_{\rm d}$). In contrast,
the (2,2) and (2,3) elements in Pattern S1 have
the same power of $\varepsilon_{\rm u,d}$, implying that
the effect of $h_{{\rm u}22}/h_{{\rm u}23}$ 
(or $h_{{\rm d}22}/h_{{\rm d}23}$) might be non-negligible.
Hence we introduce two factors $\omega_{\rm u,d}$ of $O(1)$
for the (2,2) elements of $M_{\rm u,d}$ in Pattern S1, so as
to signify the underlying difference between 
$h_{{\rm u}22}$ (or $h_{{\rm d}22}$) and $h_{{\rm u}23}$
(or $h_{{\rm d}23}$). It is reasonable to
take $h_{{\rm u}ij}/h_{{\rm u}33}=1$ and 
$h_{{\rm d}ij}/h_{{\rm d}33}=1$ for other elements of Pattern S1.
Phenomenologically we find that such a free parameter ($\omega_{\rm u}$
or $\omega_{\rm d}$) may be crucial for us to properly reproduce
the quark mass eigenvalues from $M_{\rm u}$ or $M_{\rm d}$ \cite{stringm22}.
For the same reason, we introduce factors $\omega_{\rm u,d}$ for 
the (2,2) elements of all the above-derived quark mass matrices 
except $M_{\rm u}$ of Pattern A1, where the (1,2) and (2,2) elements
have got different powers of $\varepsilon_{\rm u}$. 
Furthermore, we assume all parameters in $M_{\rm u,d}$ 
(i.e., $c_{\rm u,d}$, $\varepsilon_{\rm u,d}$ and $\omega_{\rm u,d}$) to be 
real.
Our analytical results for quark mass eigenvalues, in leading order
approximations, are listed in Table 2.
\begin{table}
\caption{Quark mass eigenvalues derived from four types of quark mass matrices.}
\vspace{-0.2cm}
\begin{center}
\begin{tabular}{ccccccccc} \\ \hline\hline \\
Pattern    & ~~~~~ & $m_t$ & $m_c$         & $m_u$         & ~~~~~
         & $m_b$        & $m_s$         & $m_d$  \\ \\ \hline \\
S1      
&& $c_{\rm u}$  & $\varepsilon_{\rm u}^2 \omega_{\rm u} c_{\rm u}$
& $\displaystyle\frac{\varepsilon^4_{\rm u}}{\omega_{\rm u}} c_{\rm u}$ 
&& $c_{\rm d}$  & $\varepsilon^2_{\rm d} \omega_{\rm d} c_{\rm d}$
& $\displaystyle\frac{\varepsilon^4_{\rm d}}{\omega_{\rm d}} c_{\rm d}$ \\ \\
S2      
&& $c_{\rm u}$  & $\varepsilon_{\rm u}^3 \omega_{\rm u} c_{\rm u}$
& $\displaystyle\frac{\varepsilon^5_{\rm u}}{\omega_{\rm u}} c_{\rm u}$ 
&& $c_{\rm d}$  & $\varepsilon^3_{\rm d} \omega_{\rm d} c_{\rm d}$
& $\displaystyle\frac{\varepsilon^5_{\rm d}}{\omega_{\rm d}} c_{\rm d}$ \\ \\
A1      
&& $c_{\rm u}$  & $\varepsilon_{\rm u}^2 c_{\rm u}$
& $\varepsilon^4_{\rm u} c_{\rm u}$ 
&& $c_{\rm d}$  & $\varepsilon^3_{\rm d} \omega_{\rm d} c_{\rm d}$
& $\displaystyle\frac{\varepsilon^5_{\rm d}}{\omega_{\rm d}} c_{\rm d}$ \\ \\
A2      
&& $c_{\rm u}$  & $\varepsilon_{\rm u}^3 \omega_{\rm u} c_{\rm u}$
& $\varepsilon^6_{\rm u} c_{\rm u}$ 
&& $c_{\rm d}$  & $\varepsilon^3_{\rm d} \omega_{\rm d} c_{\rm d}$
& $\displaystyle\frac{\varepsilon^5_{\rm d}}{\omega_{\rm d}} c_{\rm d}$ \\ \\
\hline\hline
\end{tabular}
\end{center}
\end{table}

To calculate the mixing matrix of quark flavors, 
we need introduce a simple phase matrix
\begin{equation}
P \; = \; \left ( \matrix{
1       & 0     & 0 \cr
0       & e^{{\rm i} \phi}      & 0 \cr
0       & 0     & e^{{\rm i} \phi} \cr } \right ) \; ,
\end{equation}
where $\phi$ denotes the possible phase difference between $M_{\rm u}$
and $M_{\rm d}$. 
Such a $CP$-violating phase may arise from the 
dynamical details of our fermion mass generation mechanism, 
e.g., the background antisymmetric tensors in orbifold models or 
imaginary VEVs of $\theta$. 
Phenomenologically
the existence of $\phi$ is necessary for the {\it Ansatz} to properly 
reproduce the Cabibbo angle and $CP$ violation. 
The flavor mixing matrix, i.e., the Cabibbo-Kobayashi-Maskawa (CKM) matrix, 
is given by
\begin{equation}
V \; \equiv \; O^{\rm T}_{\rm u} P O_{\rm d} \; ,
\end{equation}
where $O_{\rm u}$ and $O_{\rm d}$ are the orthogonal matrices diagonalizing
$M_{\rm u}$ and $M_{\rm d}$, respectively. Due to the 
hierarchy of quark masses, it is easy to show that all the above
four patterns lead to
$|V_{ud}| \approx |V_{cs}| \approx |V_{tb}| \approx 1$
in leading order approximations. In addition, we can reproduce the
results $|V_{us}| \approx |V_{cd}|$ and $|V_{cb}| \approx |V_{ts}|$, which
hold at both low and high energy scales \cite{XingJPG} 
(as indicated by current data and unitarity of $V$ \cite{PDG96,XingNPB}). 
The analytical results for $|V_{us}|$, $|V_{cb}|$, $|V_{ub}/V_{cb}|$
and $|V_{td}/V_{ts}|$ are listed in Table 3. 
\begin{table}
\caption{Leading order results for CKM matrix elements $|V_{us}|$, $|V_{cb}|$,
$|V_{ub}|$ and $|V_{td}|$.}
\vspace{-0.2cm}
\begin{center}
\begin{tabular}{cccccc} \\ \hline\hline \\
Pattern    & ~~~~~ & ~~~~~~~ $|V_{us}|$ ~~~~~~~    & ~~~~ $|V_{cb}|$ ~~~~  
& ~~~~~ $\displaystyle \left |\frac{V_{ub}}{V_{cb}} \right |$ ~~~~~     
& ~~~~~ $\displaystyle \left |\frac{V_{td}}{V_{ts}} \right |$  ~~~~~ \\ \\ 
\hline \\
S1      
&& $\displaystyle \left | \frac{\varepsilon_{\rm u}}{\omega_{\rm u}} e^{{\rm 
i}\phi}
- \frac{\varepsilon_{\rm d}}{\omega_{\rm d}} \right |$
& $\displaystyle \left |\varepsilon^2_{\rm u} - \varepsilon^2_{\rm d} \right |$
& $\displaystyle \left |\frac{\varepsilon_{\rm u}}{\omega_{\rm u}} \right |$
& $\displaystyle \left |\frac{\varepsilon_{\rm d}}{\omega_{\rm d}} \right |$ \\ 
\\
S2      
&& $\displaystyle \left | \frac{\varepsilon_{\rm u}}{\omega_{\rm u}} e^{{\rm 
i}\phi}
- \frac{\varepsilon_{\rm d}}{\omega_{\rm d}} \right |$
& $\displaystyle \left |\varepsilon^3_{\rm u} - \varepsilon^3_{\rm d} \right |$
& $\displaystyle \left |\frac{\varepsilon_{\rm u}}{\omega_{\rm u}} \right |$
& $\displaystyle \left |\frac{\varepsilon_{\rm d}}{\omega_{\rm d}} \right |$ \\ 
\\
A1      
&& $\displaystyle \left | \varepsilon_{\rm u} e^{{\rm i}\phi}
- \frac{\varepsilon_{\rm d}}{\omega_{\rm d}} \right |$
& $\displaystyle \left | \varepsilon^3_{\rm d} \right |$
& $\displaystyle \left | \varepsilon_{\rm u} \right |$
& $\displaystyle \left |\frac{\varepsilon_{\rm d}}{\omega_{\rm d}} \right |$ \\ 
\\
A2      
&& $\displaystyle \left | \frac{\varepsilon_{\rm d}}{\omega_{\rm d}} \right |$
& $\displaystyle \left | \varepsilon^3_{\rm d} \right |$
& $\displaystyle \left |\frac{\varepsilon^3_{\rm u}}{\omega_{\rm d} 
\varepsilon^3_{\rm d}} \right |$
& $\displaystyle \left |\frac{\varepsilon_{\rm d}}{\omega_{\rm d}} e^{{\rm 
i}\phi}
+ \frac{\varepsilon^3_{\rm u}}{\varepsilon^3_{\rm d}} \right |$ \\ \\
\hline\hline
\end{tabular}
\end{center}
\end{table}

\section{Quark mixings at the weak scale}
\setcounter{equation}{0}

Now let us confront the above-obtained flavor mixing matrix $V$  
with low-energy experimental data, in order to phenomenologically ``justify''
the string-inspired quark mass matrices $M_{\rm u,d}$.
For this purpose, we have to run the elements of $V$ from the string scale 
$M_S$ ($\sim 10^{17}$ GeV) to the weak scale $M_Z$ ($\sim 10^2$ GeV). 
Theoretically it is instructive to assume the minimal supersymmetric standard
model (MSSM) below $M_S$ \cite{stringm22}. Then one can apply the 
renormalization-group 
equations to $M_{\rm u,d}$ and $V$ in the framework of the MSSM.

\begin{center}
{\large\bf A. ~ Scale-independent results}
\end{center}

The one-loop renormalization group equations for quark
mass ratios and flavor mixing matrix elements have been explicitly 
presented in Ref. \cite{Babu93}.
In view of the hierarchy of Yukawa couplings and quark mixing angles, one can 
make reliable analytical approximations for the relevant evolution equations 
by keeping only the leading terms. It is straightforward to find that 
the running effects of $m_u/m_c$ and $m_d/m_s$ are negligibly small.
The evolutions of $|V_{us}|$ and $|V_{cd}|$ involve 
the second-family Yukawa couplings, thus they can be safely
neglected. In addition, the diagonal elements of $V$ have negligible 
evolutions with energy. Only $|V_{ub}|$, $|V_{cb}|$, $|V_{td}|$ 
and $|V_{ts}|$ may have significant renormalization-group effects, and their
running behaviors are indeed identical in leading order approximations 
\cite{Babu93}.

Taking the above points into account, we find that 
$\varepsilon_{\rm u}/\omega_{\rm u} \approx \sqrt{m_u/m_c}$ and
$\varepsilon_{\rm d}/\omega_{\rm d} \approx \sqrt{m_d/m_s}$ in Pattern S1 are
approximately scale-independent. So are the parameters $\varepsilon_{\rm u}
/\omega_{\rm u} \approx \sqrt{m_u/m_c}$ and $\varepsilon_{\rm d}/\omega_{\rm d}
\approx \sqrt{m_d/m_s}$ in Pattern S2; $\varepsilon_{\rm u} \approx 
\sqrt{m_u/m_c}$ and $\varepsilon_{\rm d}/\omega_{\rm d} \approx \sqrt{m_d/m_s}$
in Pattern A1; $\varepsilon^3_{\rm u}/\omega_{\rm u} \approx m_u/m_c$ and
$\varepsilon_{\rm d}/\omega_{\rm d} \approx \sqrt{m_d/m_s}$ in Pattern A2.
Consequently some scale-independent 
results for quark mixings at the weak scale $M_Z$ can be straightforwardly 
obtained, as listed in Table 4.
\begin{table}
\caption{Leading order results for $|V_{us}|$, $|V_{ub}/V_{cb}|$ and
$|V_{td}/V_{ts}|$ at $M_Z$.}
\vspace{-0.2cm}
\begin{center}
\begin{tabular}{ccccc} \\ \hline\hline \\
Pattern    & ~~~~~ & ~~~~~~~ $|V_{us}|$ ~~~~~~~    
& ~~~~~ $\displaystyle \left |\frac{V_{ub}}{V_{cb}} \right |$ ~~~~~     
& ~~~~~ $\displaystyle \left |\frac{V_{td}}{V_{ts}} \right |$  ~~~~~ \\ \\ 
\hline \\
S1      
&& $\displaystyle \left | \sqrt{\frac{m_u}{m_c}} e^{{\rm i}\phi}
- \sqrt{\frac{m_d}{m_s}} \right |$
& $\displaystyle \sqrt{\frac{m_u}{m_c}}$
& $\displaystyle \sqrt{\frac{m_d}{m_s}}$ \\ \\
S2      
&& $\displaystyle \left | \sqrt{\frac{m_u}{m_c}} e^{{\rm i}\phi}
- \sqrt{\frac{m_d}{m_s}} \right |$
& $\displaystyle \sqrt{\frac{m_u}{m_c}}$
& $\displaystyle \sqrt{\frac{m_d}{m_s}}$ \\ \\
A1      
&& $\displaystyle \left | \sqrt{\frac{m_u}{m_c}} e^{{\rm i}\phi}
- \sqrt{\frac{m_d}{m_s}} \right |$
& $\displaystyle \sqrt{\frac{m_u}{m_c}}$
& $\displaystyle \sqrt{\frac{m_d}{m_s}}$ \\ \\
A2      
&& $\displaystyle \sqrt{\frac{m_d}{m_s}}$
& ---
& --- \\ \\
\hline\hline
\end{tabular}
\end{center}
\end{table}

Some discussions about Table 4 are in order:

(a) For Pattern S1, S2 or A1, the $CP$-violating phase $\phi$ can be determined
from the present data on $|V_{us}|$:
\begin{equation}
\phi \; \approx \; \arccos \left [ \frac{1}{2} \sqrt{ \frac{m_c m_s}{m_u m_d} }
\left ( \frac{m_u}{m_c} + \frac{m_d}{m_s} - |V_{us}|^2 \right ) \right ] \; .
\end{equation}
By use of $|V_{us}|_{\rm exp} =0.2205 \pm 0.0018$ \cite{PDG96}, 
$m_s/m_d =19.3\pm 0.9$ \cite{Leutwyler96} and $10^{-3} \leq m_u/m_c \leq 
10^{-2}$
\cite{PDG96}, we calculate $\phi$ and find an allowed region for its value in 
the 
first quadrant: $75^0 \leq \phi \leq 85^0$. Current data on $CP$ violation
in $K^0$-$\bar{K}^0$ mixing (i.e., $\epsilon^{~}_K$) 
\cite{PDG96} have excluded the possibility of $\phi < 0$.

(b) For Pattern A2, we get $|V_{us}| \approx \sqrt{m_d/m_s} \approx 0.228 \pm 
0.005$, in agreement with the present experimental value 
$|V_{us}|_{\rm exp}=0.2205 \pm 0.0018$. In this
pattern, the $CP$-violating phase is in principle determinable from 
$|V_{td}/V_{ts}|$.

(c) The instructive relations 
$|V_{ub}/V_{cb}| \approx \sqrt{m_u/m_c}$ and 
$|V_{td}/V_{ts}| \approx \sqrt{m_d/m_s}$ hold approximately in
Patterns S1, S2 and A1. They result from the texture zeros of (1,1), (1,3)
and (3,1) elements for both up and down mass matrices 
\cite{Fritzsch78,FritzschXing95,Hall93}. Typically taking
$m_u = 5.1 \pm 0.9$ MeV and $m_c = 1.3$ GeV at the scale $\mu =1$ GeV 
\cite{Gasser82}, 
we predict $|V_{ub}/V_{cb}| = 0.063 \pm 0.005$ and $|V_{td}/V_{ts}| = 0.228 \pm 
0.005$.
In comparison, current measurements have given $|V_{ub}/V_{cb}|_{\rm exp}
=0.08 \pm 0.02$ \cite{PDG96} and constrained the ratio of $|V_{td}|$ to 
$|V_{ts}|$
to be in the region 0.15 -- 0.34 \cite{Ali96}.

\begin{center}
{\large\bf B. ~ Scale-dependent results}
\end{center}

For all four patterns of quark mass matrices, the analytical expressions of
$|V_{cb}|$ and $|V_{ts}|$ obtained in Table 3 will be altered due to 
non-negligible running effects from the string scale $\mu =M_S$ to the 
weak scale $\mu =M_Z$. The results for $|V_{ub}/V_{cb}|$ and $|V_{td}/V_{ts}|$
in Pattern A2 are also sensitive to the renormalization-group effects, since
they depend strongly upon the mass ratios $m_c/m_t$ and $m_s/m_b$. The
relevant evolution functions can be defined by \cite{Xing96}
\begin{equation}
\xi_{t,b} \; = \; \exp \left [ -\frac{1}{16\pi^2} 
\int^{\ln (M_S/M_Z)} _0 f^2_{t,b}(\chi) ~ {\rm d}\chi \right ] \; ,
\end{equation}
where $\chi \equiv \ln (\mu /M_Z)$; and $f_t$ and $f_b$ are Yukawa
coupling eigenvalues of the top and bottom quarks, respectively.
A good approximation is that the third-family Yukawa couplings 
of quarks (and charged leptons), together with the gauge couplings, 
play the dominant roles in the renormalization-group equations \cite{Babu93}.
Then the magnitudes of $\xi_t$ and $\xi_b$ can be evaluated for arbitrary
$\tan\beta$ (the ratio of Higgs vacuum expectation values in the MSSM)
from $M_S$ to $M_Z$, and the numerical results
have been given in \cite{Xing96}. In terms of $\xi_t$ and $\xi_b$,
three key evolution relations based on the MSSM read \cite{Xing96}:
\begin{eqnarray}
\left . \frac{m_s}{m_b} \right |_{M_Z} & = & \frac{1}{\xi_t ~ \xi^3_b} ~
\left . \frac{m_s}{m_b} \right |_{M_S} \; , \nonumber \\
\left . \frac{m_c}{m_t} \right |_{M_Z} & = & \frac{1}{\xi^3_t ~ \xi_b} ~ 
\left . \frac{m_c}{m_t} \right |_{M_S}  \; , \nonumber \\
\left |V_{cb} \right |_{M_Z} & = & \frac{1}{\xi_t ~ \xi_b} ~ 
\left |V_{cb} \right |_{M_S} \; .
\end{eqnarray}
The running behaviors of $|V_{ub}|$, $|V_{td}|$ and $|V_{ts}|$ are identical
to that of $|V_{cb}|$. By use of Eq. (3.3), we are able to 
obtain the renormalized expressions of $|V_{cb}|$ at $M_Z$, as listed in Table 
5.
\begin{table}
\caption{Leading order results for $|V_{cb}|$ at the weak scale $M_Z$.}
\vspace{-0.2cm}
\begin{center}
\begin{tabular}{ccc} \\ \hline\hline \\
Pattern    & ~~~~~ & ~~~~~~~ Renormalized $|V_{cb}|$ ~~~~~~~    \\ \\ \hline \\
S1      
&& $\displaystyle \left ( \xi_t^{-1/3} \xi_b \right ) \left (\frac{m_d}{m_s}
\right )^{1/3} \left (\frac{m_s}{m_b} \right )^{2/3} ~ - ~ \left ( \xi_t 
\xi_b^{-1/3} \right ) \left (\frac{m_u}{m_c} \right )^{1/3} \left (\frac{m_c}
{m_t} \right )^{2/3}$  \\ \\
S2
&& $\displaystyle \left ( \xi_t^{-1/4} \xi_b^{5/4} \right ) \left 
(\frac{m_d}{m_s}
\right )^{3/8} \left (\frac{m_s}{m_b} \right )^{3/4} ~ - ~ \left ( \xi_t^{5/4} 
\xi_b^{-1/4} \right ) \left (\frac{m_u}{m_c} \right )^{3/8} \left (\frac{m_c}
{m_t} \right )^{3/4}$  \\ \\
A1      
&& $\displaystyle \left ( \xi_t^{-1/4} \xi_b^{5/4} \right ) \left 
(\frac{m_d}{m_s}
\right )^{3/8} \left (\frac{m_s}{m_b} \right )^{3/4}$ \\ \\
A2
&& $\displaystyle \left ( \xi_t^{-1/4} \xi_b^{5/4} \right ) \left 
(\frac{m_d}{m_s}
\right )^{3/8} \left (\frac{m_s}{m_b} \right )^{3/4}$
\\ \\
\hline\hline
\end{tabular}
\end{center}
\end{table}

At the weak scale $M_Z$, the ratios $|V_{ub}/V_{cb}|$ and $|V_{td}/V_{ts}|$
obtained in Pattern A2 can be renormalized as
\begin{equation}
\left | \frac{V_{ub}}{V_{cb}} \right | \; \approx \; \left ( \xi^{1/2}_t
\xi^{-5/2}_b \right ) \left ( \frac{m_u}{m_c} \right )^{1/2} 
\left ( \frac{m_c}{m_t} \right )^{1/2} \left ( \frac{m_s}{m_b} \right )^{-1} \; 
\end{equation}
and
\begin{equation}
\left | \frac{V_{td}}{V_{ts}} \right | \; \approx \; \left | \left ( 
\frac{m_d}{m_s}
\right )^{1/2} e^{{\rm i}\phi} ~ + ~ \left ( \xi^{3/4}_t \xi^{-7/4}_b \right )
\left ( \frac{m_u}{m_c} \right )^{1/2} \left ( \frac{m_d}{m_s} \right )^{-3/8}
\left ( \frac{m_c}{m_t} \right )^{1/2} \left ( \frac{m_s}{m_b} \right )^{-3/4}
\right | \; ,
\end{equation}
respectively. We observe that these two analytical expressions, although their
quantitative results could be compatible with current data, are qualitatively
less interesting than
those simpler ones derived from Patterns S1, S2 and A1.

For the purpose of illustration, we make an estimation of 
the above scale-dependent $|V_{cb}|$, $|V_{ub}/V_{cb}|$ and 
$|V_{td}/V_{ts}|$. We take $m_u/m_c = 0.004$,
$m_c/m_t = 0.005$, $m_d/m_s =0.05$ and $m_s/m_b =0.035$ at $M_Z$ typically
\cite{PDG96,Gasser82}.
With the help of the numerical results of $\xi_t$ and $\xi_b$ obtained in Ref. 
\cite{Xing96}, we plot the renormalized $|V_{cb}|$ as a function of $\tan\beta$ 
in Fig. 1 for all
four patterns of quark mass matrices; and illustrate the renormalized 
$|V_{ub}/V_{cb}|$ and $|V_{td}/V_{ts}|$ changing with $\tan\beta$ in Fig. 2 for
Pattern A2. Some comments on Figs. 1 and 2 are in order.

(1) The uncertainty of $|V_{cb}|$, arising from the unknown value of 
$\tan\beta$,
may be as large as 0.01. This error is obviously larger than the experimental
error in determining $|V_{cb}|$ (e.g., $|V_{cb}|_{\rm exp} = 0.039 \pm 0.003$
\cite{Ali96}). The similar problem exists for the renormalized $|V_{ub}/V_{cb}|$
and $|V_{td}/V_{ts}|$ in Pattern A2. Hence it is difficult, even impossible,
to numerically ``justify'' the proposed mass matrix patterns through their 
consequences
on $|V_{cb}|$, before the value of $\tan\beta$ in the MSSM can be reliably fixed
somewhere else.

(2) With the inputs taken above and appropriate values of $\tan\beta$, we find 
that
only Pattern S1 is likely to properly reproduce $|V_{cb}|$ 
at the weak scale \cite{stringm22}.
Of course the other three patterns could also lead to proper results of 
$|V_{cb}|$,
if one changes the input values of quark mass ratios and takes their large 
errors
into account. The interesting point is that the value of the renormalized 
$|V_{cb}|$ in 
Pattern S1 is remarkably larger than those in Patterns S2, A1 and A2. Therefore
it is in principle possible to distinguish Pattern S1 from the others through
the window of $|V_{cb}|$. Numerically, however, it is extremely difficult
to distinguish between Patterns S2 and A1.

(3) Pattern A2 can be ruled out if it fails in reproducing the experimental
values of $|V_{cb}|$ and $|V_{ub}/V_{cb}|$, simultaneously, with suitable 
inputs of quark mass ratios. In view of the illustrative Figs. 1 and 2, we find
that both $|V_{cb}|$ and $|V_{ub}/V_{cb}|$ obtained from Pattern A2 are not
favored by current data for all possible values of $\tan\beta$. Indeed this 
pattern is less attractive than the others due to its complicated analytical 
consequences
on $|V_{ub}/V_{cb}|$ and $|V_{td}/V_{ts}|$, as we have pointed out before.

\section{Concluding remarks}

We have studied a few typical patterns of quark mass matrices within the 
framework of orbifold models.
Two of the five RRR-type mass matrices have been derived, but it is difficult 
to obtain the others from $Z_6-$II orbifold models.
One of the main difficulties is that one cannot generate several 
hierarchies in a quark mass matrix by only using selection rules of orbifold 
models, 
because the number of twisted sectors is limited.
This is a remarkable difference of our approach from that with gauge symmetries, 
where any charge to generate more hierarchies could be assigned. 
In addition, we have derived two types of realistic quark mass matrices 
with four texture zeros and up-down structural parallelism.

For the purpose of illustration, we have confronted the obtained results of
quark flavor mixings at the string scale with low-energy data by use of the
renormalization-group equations. With the same input values of quark mass 
ratios,
we find that the four mass matrix
patterns proposed above cannot all be in good agreement with current 
experimental data.
It is worth emphasizing that some instructive
scale-independent relations like $|V_{ub}/V_{cb}| \approx \sqrt{m_u/m_c}$ and
$|V_{td}/V_{ts}| \approx \sqrt{m_d/m_s}$, 
which are favored by current experimental data, might imply a right way in
phenomenology towards the correct pattern of quark mass matrices at the 
superstring
scale. 

One can extend the analyses made in this work so as to obtain other types of 
quark
mass matrices. Also it should be interesting to apply our approach to the lepton 
mass
matrices. 

\vspace{0.5cm}
\begin{flushleft}
{\Large\bf Acknowledgments}
\end{flushleft}

One of us (ZZX) would like to thank H. Fritzsch for his helpful comments
on the string-inspired quark mass $Ans\ddot{a}tze$. He is also
grateful to A.I. Sanda for his warm hospitality and to the Japan Society for
the Promotion of Science for its financial support.

\newpage

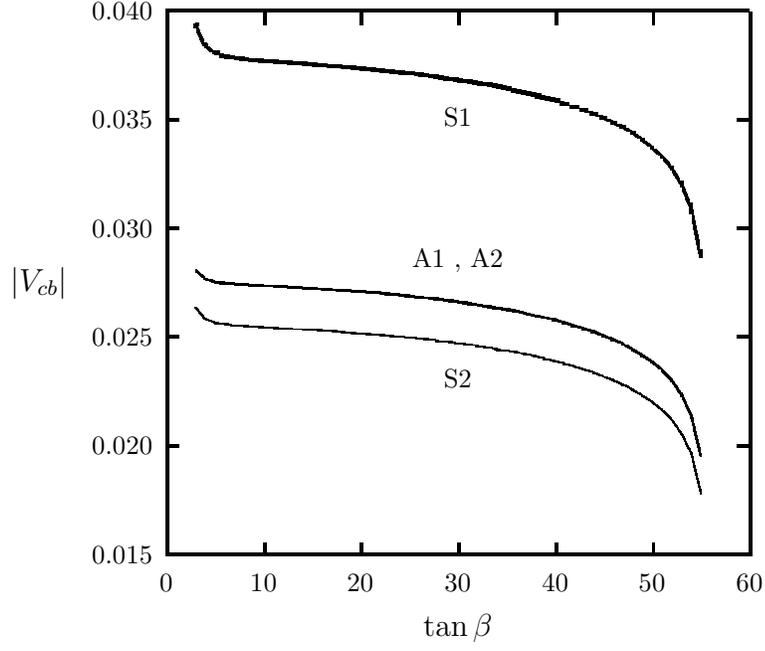
\begin{figure}
\setlength{\unitlength}{0.240900pt}
\ifx\plotpoint\undefined\newsavebox{\plotpoint}\fi
\sbox{\plotpoint}{\rule[-0.500pt]{1.000pt}{1.000pt}}%
\begin{picture}(1200,990)(-350,0)
\font\gnuplot=cmr10 at 10pt
\gnuplot
\sbox{\plotpoint}{\rule[-0.500pt]{1.000pt}{1.000pt}}%
\put(220.0,113.0){\rule[-0.500pt]{1.000pt}{205.729pt}}
\put(220.0,113.0){\rule[-0.500pt]{4.818pt}{1.000pt}}
\put(198,113){\makebox(0,0)[r]{0.015}}
\put(1116.0,113.0){\rule[-0.500pt]{4.818pt}{1.000pt}}
\put(220.0,284.0){\rule[-0.500pt]{4.818pt}{1.000pt}}
\put(198,284){\makebox(0,0)[r]{0.020}}
\put(1116.0,284.0){\rule[-0.500pt]{4.818pt}{1.000pt}}
\put(220.0,455.0){\rule[-0.500pt]{4.818pt}{1.000pt}}
\put(198,455){\makebox(0,0)[r]{0.025}}
\put(1116.0,455.0){\rule[-0.500pt]{4.818pt}{1.000pt}}
\put(220.0,625.0){\rule[-0.500pt]{4.818pt}{1.000pt}}
\put(198,625){\makebox(0,0)[r]{0.030}}
\put(1116.0,625.0){\rule[-0.500pt]{4.818pt}{1.000pt}}
\put(220.0,796.0){\rule[-0.500pt]{4.818pt}{1.000pt}}
\put(198,796){\makebox(0,0)[r]{0.035}}
\put(1116.0,796.0){\rule[-0.500pt]{4.818pt}{1.000pt}}
\put(220.0,967.0){\rule[-0.500pt]{4.818pt}{1.000pt}}
\put(198,967){\makebox(0,0)[r]{0.040}}
\put(1116.0,967.0){\rule[-0.500pt]{4.818pt}{1.000pt}}
\put(220.0,113.0){\rule[-0.500pt]{1.000pt}{4.818pt}}
\put(220,68){\makebox(0,0){0}}
\put(220.0,947.0){\rule[-0.500pt]{1.000pt}{4.818pt}}
\put(373.0,113.0){\rule[-0.500pt]{1.000pt}{4.818pt}}
\put(373,68){\makebox(0,0){10}}
\put(373.0,947.0){\rule[-0.500pt]{1.000pt}{4.818pt}}
\put(525.0,113.0){\rule[-0.500pt]{1.000pt}{4.818pt}}
\put(525,68){\makebox(0,0){20}}
\put(525.0,947.0){\rule[-0.500pt]{1.000pt}{4.818pt}}
\put(678.0,113.0){\rule[-0.500pt]{1.000pt}{4.818pt}}
\put(678,68){\makebox(0,0){30}}
\put(678.0,947.0){\rule[-0.500pt]{1.000pt}{4.818pt}}
\put(831.0,113.0){\rule[-0.500pt]{1.000pt}{4.818pt}}
\put(831,68){\makebox(0,0){40}}
\put(831.0,947.0){\rule[-0.500pt]{1.000pt}{4.818pt}}
\put(983.0,113.0){\rule[-0.500pt]{1.000pt}{4.818pt}}
\put(983,68){\makebox(0,0){50}}
\put(983.0,947.0){\rule[-0.500pt]{1.000pt}{4.818pt}}
\put(1136.0,113.0){\rule[-0.500pt]{1.000pt}{4.818pt}}
\put(1136,68){\makebox(0,0){60}}
\put(1136.0,947.0){\rule[-0.500pt]{1.000pt}{4.818pt}}
\put(220.0,113.0){\rule[-0.500pt]{220.664pt}{1.000pt}}
\put(1136.0,113.0){\rule[-0.500pt]{1.000pt}{205.729pt}}
\put(220.0,967.0){\rule[-0.500pt]{220.664pt}{1.000pt}}
\put(20,540){\makebox(0,0){$|V_{cb}|$}}
\put(678,-3){\makebox(0,0){$\tan\beta$}}
\put(678,800){\makebox(0,0){S1}}
\put(678,575){\makebox(0,0){A1 , A2}}
\put(678,390){\makebox(0,0){S2}}

\put(220.0,113.0){\rule[-0.500pt]{1.000pt}{205.729pt}}
\put(266,946){\usebox{\plotpoint}}
\multiput(267.83,935.55)(0.493,-1.122){22}{\rule{0.119pt}{2.517pt}}
\multiput(263.92,940.78)(15.000,-28.777){2}{\rule{1.000pt}{1.258pt}}
\multiput(281.00,909.68)(0.639,-0.489){14}{\rule{1.614pt}{0.118pt}}
\multiput(281.00,909.92)(11.651,-11.000){2}{\rule{0.807pt}{1.000pt}}
\multiput(296.00,898.71)(1.679,-0.424){2}{\rule{3.450pt}{0.102pt}}
\multiput(296.00,898.92)(8.839,-5.000){2}{\rule{1.725pt}{1.000pt}}
\put(312,892.42){\rule{3.614pt}{1.000pt}}
\multiput(312.00,893.92)(7.500,-3.000){2}{\rule{1.807pt}{1.000pt}}
\put(327,889.92){\rule{3.614pt}{1.000pt}}
\multiput(327.00,890.92)(7.500,-2.000){2}{\rule{1.807pt}{1.000pt}}
\put(342,887.92){\rule{3.614pt}{1.000pt}}
\multiput(342.00,888.92)(7.500,-2.000){2}{\rule{1.807pt}{1.000pt}}
\put(357,886.42){\rule{3.854pt}{1.000pt}}
\multiput(357.00,886.92)(8.000,-1.000){2}{\rule{1.927pt}{1.000pt}}
\put(373,885.42){\rule{3.614pt}{1.000pt}}
\multiput(373.00,885.92)(7.500,-1.000){2}{\rule{1.807pt}{1.000pt}}
\put(388,884.42){\rule{3.614pt}{1.000pt}}
\multiput(388.00,884.92)(7.500,-1.000){2}{\rule{1.807pt}{1.000pt}}
\put(403,883.42){\rule{3.614pt}{1.000pt}}
\multiput(403.00,883.92)(7.500,-1.000){2}{\rule{1.807pt}{1.000pt}}
\put(418,882.42){\rule{3.854pt}{1.000pt}}
\multiput(418.00,882.92)(8.000,-1.000){2}{\rule{1.927pt}{1.000pt}}
\put(434,881.42){\rule{3.614pt}{1.000pt}}
\multiput(434.00,881.92)(7.500,-1.000){2}{\rule{1.807pt}{1.000pt}}
\put(449,879.92){\rule{3.614pt}{1.000pt}}
\multiput(449.00,880.92)(7.500,-2.000){2}{\rule{1.807pt}{1.000pt}}
\put(464,878.42){\rule{3.854pt}{1.000pt}}
\multiput(464.00,878.92)(8.000,-1.000){2}{\rule{1.927pt}{1.000pt}}
\put(480,877.42){\rule{3.614pt}{1.000pt}}
\multiput(480.00,877.92)(7.500,-1.000){2}{\rule{1.807pt}{1.000pt}}
\put(495,876.42){\rule{3.614pt}{1.000pt}}
\multiput(495.00,876.92)(7.500,-1.000){2}{\rule{1.807pt}{1.000pt}}
\put(510,874.92){\rule{3.614pt}{1.000pt}}
\multiput(510.00,875.92)(7.500,-2.000){2}{\rule{1.807pt}{1.000pt}}
\put(525,873.42){\rule{3.854pt}{1.000pt}}
\multiput(525.00,873.92)(8.000,-1.000){2}{\rule{1.927pt}{1.000pt}}
\put(541,871.92){\rule{3.614pt}{1.000pt}}
\multiput(541.00,872.92)(7.500,-2.000){2}{\rule{1.807pt}{1.000pt}}
\put(556,870.42){\rule{3.614pt}{1.000pt}}
\multiput(556.00,870.92)(7.500,-1.000){2}{\rule{1.807pt}{1.000pt}}
\put(571,868.92){\rule{3.614pt}{1.000pt}}
\multiput(571.00,869.92)(7.500,-2.000){2}{\rule{1.807pt}{1.000pt}}
\put(586,867.42){\rule{3.854pt}{1.000pt}}
\multiput(586.00,867.92)(8.000,-1.000){2}{\rule{1.927pt}{1.000pt}}
\put(602,865.92){\rule{3.614pt}{1.000pt}}
\multiput(602.00,866.92)(7.500,-2.000){2}{\rule{1.807pt}{1.000pt}}
\put(617,863.92){\rule{3.614pt}{1.000pt}}
\multiput(617.00,864.92)(7.500,-2.000){2}{\rule{1.807pt}{1.000pt}}
\put(632,861.92){\rule{3.614pt}{1.000pt}}
\multiput(632.00,862.92)(7.500,-2.000){2}{\rule{1.807pt}{1.000pt}}
\put(647,859.42){\rule{3.854pt}{1.000pt}}
\multiput(647.00,860.92)(8.000,-3.000){2}{\rule{1.927pt}{1.000pt}}
\put(663,856.92){\rule{3.614pt}{1.000pt}}
\multiput(663.00,857.92)(7.500,-2.000){2}{\rule{1.807pt}{1.000pt}}
\put(678,854.92){\rule{3.614pt}{1.000pt}}
\multiput(678.00,855.92)(7.500,-2.000){2}{\rule{1.807pt}{1.000pt}}
\put(693,852.42){\rule{3.854pt}{1.000pt}}
\multiput(693.00,853.92)(8.000,-3.000){2}{\rule{1.927pt}{1.000pt}}
\put(709,849.92){\rule{3.614pt}{1.000pt}}
\multiput(709.00,850.92)(7.500,-2.000){2}{\rule{1.807pt}{1.000pt}}
\put(724,847.42){\rule{3.614pt}{1.000pt}}
\multiput(724.00,848.92)(7.500,-3.000){2}{\rule{1.807pt}{1.000pt}}
\put(739,844.42){\rule{3.614pt}{1.000pt}}
\multiput(739.00,845.92)(7.500,-3.000){2}{\rule{1.807pt}{1.000pt}}
\put(754,840.92){\rule{3.854pt}{1.000pt}}
\multiput(754.00,842.92)(8.000,-4.000){2}{\rule{1.927pt}{1.000pt}}
\put(770,837.42){\rule{3.614pt}{1.000pt}}
\multiput(770.00,838.92)(7.500,-3.000){2}{\rule{1.807pt}{1.000pt}}
\put(785,833.92){\rule{3.614pt}{1.000pt}}
\multiput(785.00,835.92)(7.500,-4.000){2}{\rule{1.807pt}{1.000pt}}
\put(800,829.92){\rule{3.614pt}{1.000pt}}
\multiput(800.00,831.92)(7.500,-4.000){2}{\rule{1.807pt}{1.000pt}}
\put(815,825.92){\rule{3.854pt}{1.000pt}}
\multiput(815.00,827.92)(8.000,-4.000){2}{\rule{1.927pt}{1.000pt}}
\multiput(831.00,823.71)(1.509,-0.424){2}{\rule{3.250pt}{0.102pt}}
\multiput(831.00,823.92)(8.254,-5.000){2}{\rule{1.625pt}{1.000pt}}
\multiput(846.00,818.71)(1.509,-0.424){2}{\rule{3.250pt}{0.102pt}}
\multiput(846.00,818.92)(8.254,-5.000){2}{\rule{1.625pt}{1.000pt}}
\multiput(861.00,813.71)(1.509,-0.424){2}{\rule{3.250pt}{0.102pt}}
\multiput(861.00,813.92)(8.254,-5.000){2}{\rule{1.625pt}{1.000pt}}
\multiput(876.00,808.69)(1.298,-0.462){4}{\rule{2.917pt}{0.111pt}}
\multiput(876.00,808.92)(9.946,-6.000){2}{\rule{1.458pt}{1.000pt}}
\multiput(892.00,802.69)(1.195,-0.462){4}{\rule{2.750pt}{0.111pt}}
\multiput(892.00,802.92)(9.292,-6.000){2}{\rule{1.375pt}{1.000pt}}
\multiput(907.00,796.68)(0.883,-0.481){8}{\rule{2.125pt}{0.116pt}}
\multiput(907.00,796.92)(10.589,-8.000){2}{\rule{1.063pt}{1.000pt}}
\multiput(922.00,788.69)(1.095,-0.475){6}{\rule{2.536pt}{0.114pt}}
\multiput(922.00,788.92)(10.737,-7.000){2}{\rule{1.268pt}{1.000pt}}
\multiput(938.00,781.68)(0.783,-0.485){10}{\rule{1.917pt}{0.117pt}}
\multiput(938.00,781.92)(11.022,-9.000){2}{\rule{0.958pt}{1.000pt}}
\multiput(953.00,772.68)(0.639,-0.489){14}{\rule{1.614pt}{0.118pt}}
\multiput(953.00,772.92)(11.651,-11.000){2}{\rule{0.807pt}{1.000pt}}
\multiput(968.00,761.68)(0.585,-0.491){16}{\rule{1.500pt}{0.118pt}}
\multiput(968.00,761.92)(11.887,-12.000){2}{\rule{0.750pt}{1.000pt}}
\multiput(983.00,749.68)(0.538,-0.492){20}{\rule{1.393pt}{0.119pt}}
\multiput(983.00,749.92)(13.109,-14.000){2}{\rule{0.696pt}{1.000pt}}
\multiput(1000.83,731.98)(0.493,-0.571){22}{\rule{0.119pt}{1.450pt}}
\multiput(996.92,734.99)(15.000,-14.990){2}{\rule{1.000pt}{0.725pt}}
\multiput(1015.83,712.04)(0.493,-0.812){22}{\rule{0.119pt}{1.917pt}}
\multiput(1011.92,716.02)(15.000,-21.022){2}{\rule{1.000pt}{0.958pt}}
\multiput(1030.83,684.00)(0.493,-1.190){22}{\rule{0.119pt}{2.650pt}}
\multiput(1026.92,689.50)(15.000,-30.500){2}{\rule{1.000pt}{1.325pt}}
\multiput(1045.83,637.47)(0.494,-2.499){24}{\rule{0.119pt}{5.188pt}}
\multiput(1041.92,648.23)(16.000,-68.233){2}{\rule{1.000pt}{2.594pt}}
\sbox{\plotpoint}{\rule[-0.175pt]{0.350pt}{0.350pt}}%
\put(266,501){\usebox{\plotpoint}}
\multiput(266.48,498.89)(0.502,-0.611){27}{\rule{0.121pt}{0.507pt}}
\multiput(265.27,499.95)(15.000,-16.947){2}{\rule{0.350pt}{0.254pt}}
\multiput(281.00,482.02)(1.376,-0.505){9}{\rule{0.962pt}{0.122pt}}
\multiput(281.00,482.27)(13.002,-6.000){2}{\rule{0.481pt}{0.350pt}}
\put(296,475.27){\rule{2.888pt}{0.350pt}}
\multiput(296.00,476.27)(10.007,-2.000){2}{\rule{1.444pt}{0.350pt}}
\put(312,473.27){\rule{2.712pt}{0.350pt}}
\multiput(312.00,474.27)(9.370,-2.000){2}{\rule{1.356pt}{0.350pt}}
\put(327,471.77){\rule{3.614pt}{0.350pt}}
\multiput(327.00,472.27)(7.500,-1.000){2}{\rule{1.807pt}{0.350pt}}
\put(342,470.77){\rule{3.614pt}{0.350pt}}
\multiput(342.00,471.27)(7.500,-1.000){2}{\rule{1.807pt}{0.350pt}}
\put(357,469.77){\rule{3.854pt}{0.350pt}}
\multiput(357.00,470.27)(8.000,-1.000){2}{\rule{1.927pt}{0.350pt}}
\put(373,468.77){\rule{3.614pt}{0.350pt}}
\multiput(373.00,469.27)(7.500,-1.000){2}{\rule{1.807pt}{0.350pt}}
\put(388,467.77){\rule{3.614pt}{0.350pt}}
\multiput(388.00,468.27)(7.500,-1.000){2}{\rule{1.807pt}{0.350pt}}
\put(403,466.77){\rule{3.614pt}{0.350pt}}
\multiput(403.00,467.27)(7.500,-1.000){2}{\rule{1.807pt}{0.350pt}}
\put(434,465.77){\rule{3.614pt}{0.350pt}}
\multiput(434.00,466.27)(7.500,-1.000){2}{\rule{1.807pt}{0.350pt}}
\put(449,464.77){\rule{3.614pt}{0.350pt}}
\multiput(449.00,465.27)(7.500,-1.000){2}{\rule{1.807pt}{0.350pt}}
\put(464,463.77){\rule{3.854pt}{0.350pt}}
\multiput(464.00,464.27)(8.000,-1.000){2}{\rule{1.927pt}{0.350pt}}
\put(480,462.77){\rule{3.614pt}{0.350pt}}
\multiput(480.00,463.27)(7.500,-1.000){2}{\rule{1.807pt}{0.350pt}}
\put(495,461.27){\rule{2.712pt}{0.350pt}}
\multiput(495.00,462.27)(9.370,-2.000){2}{\rule{1.356pt}{0.350pt}}
\put(510,459.77){\rule{3.614pt}{0.350pt}}
\multiput(510.00,460.27)(7.500,-1.000){2}{\rule{1.807pt}{0.350pt}}
\put(525,458.77){\rule{3.854pt}{0.350pt}}
\multiput(525.00,459.27)(8.000,-1.000){2}{\rule{1.927pt}{0.350pt}}
\put(541,457.77){\rule{3.614pt}{0.350pt}}
\multiput(541.00,458.27)(7.500,-1.000){2}{\rule{1.807pt}{0.350pt}}
\put(556,456.77){\rule{3.614pt}{0.350pt}}
\multiput(556.00,457.27)(7.500,-1.000){2}{\rule{1.807pt}{0.350pt}}
\put(571,455.27){\rule{2.712pt}{0.350pt}}
\multiput(571.00,456.27)(9.370,-2.000){2}{\rule{1.356pt}{0.350pt}}
\put(586,453.77){\rule{3.854pt}{0.350pt}}
\multiput(586.00,454.27)(8.000,-1.000){2}{\rule{1.927pt}{0.350pt}}
\put(602,452.27){\rule{2.712pt}{0.350pt}}
\multiput(602.00,453.27)(9.370,-2.000){2}{\rule{1.356pt}{0.350pt}}
\put(617,450.77){\rule{3.614pt}{0.350pt}}
\multiput(617.00,451.27)(7.500,-1.000){2}{\rule{1.807pt}{0.350pt}}
\multiput(632.00,450.02)(3.686,-0.516){3}{\rule{1.837pt}{0.124pt}}
\multiput(632.00,450.27)(11.186,-3.000){2}{\rule{0.919pt}{0.350pt}}
\put(647,446.77){\rule{3.854pt}{0.350pt}}
\multiput(647.00,447.27)(8.000,-1.000){2}{\rule{1.927pt}{0.350pt}}
\put(663,445.27){\rule{2.712pt}{0.350pt}}
\multiput(663.00,446.27)(9.370,-2.000){2}{\rule{1.356pt}{0.350pt}}
\put(678,443.27){\rule{2.712pt}{0.350pt}}
\multiput(678.00,444.27)(9.370,-2.000){2}{\rule{1.356pt}{0.350pt}}
\multiput(693.00,442.02)(3.944,-0.516){3}{\rule{1.954pt}{0.124pt}}
\multiput(693.00,442.27)(11.944,-3.000){2}{\rule{0.977pt}{0.350pt}}
\put(709,438.27){\rule{2.712pt}{0.350pt}}
\multiput(709.00,439.27)(9.370,-2.000){2}{\rule{1.356pt}{0.350pt}}
\multiput(724.00,437.02)(3.686,-0.516){3}{\rule{1.837pt}{0.124pt}}
\multiput(724.00,437.27)(11.186,-3.000){2}{\rule{0.919pt}{0.350pt}}
\put(739,433.27){\rule{2.712pt}{0.350pt}}
\multiput(739.00,434.27)(9.370,-2.000){2}{\rule{1.356pt}{0.350pt}}
\multiput(754.00,432.02)(3.944,-0.516){3}{\rule{1.954pt}{0.124pt}}
\multiput(754.00,432.27)(11.944,-3.000){2}{\rule{0.977pt}{0.350pt}}
\multiput(770.00,429.02)(3.686,-0.516){3}{\rule{1.837pt}{0.124pt}}
\multiput(770.00,429.27)(11.186,-3.000){2}{\rule{0.919pt}{0.350pt}}
\multiput(785.00,426.02)(3.686,-0.516){3}{\rule{1.837pt}{0.124pt}}
\multiput(785.00,426.27)(11.186,-3.000){2}{\rule{0.919pt}{0.350pt}}
\multiput(800.00,423.02)(2.297,-0.509){5}{\rule{1.400pt}{0.123pt}}
\multiput(800.00,423.27)(12.094,-4.000){2}{\rule{0.700pt}{0.350pt}}
\multiput(815.00,419.02)(2.456,-0.509){5}{\rule{1.487pt}{0.123pt}}
\multiput(815.00,419.27)(12.913,-4.000){2}{\rule{0.744pt}{0.350pt}}
\multiput(831.00,415.02)(2.297,-0.509){5}{\rule{1.400pt}{0.123pt}}
\multiput(831.00,415.27)(12.094,-4.000){2}{\rule{0.700pt}{0.350pt}}
\multiput(846.00,411.02)(2.297,-0.509){5}{\rule{1.400pt}{0.123pt}}
\multiput(846.00,411.27)(12.094,-4.000){2}{\rule{0.700pt}{0.350pt}}
\multiput(861.00,407.02)(1.713,-0.507){7}{\rule{1.137pt}{0.122pt}}
\multiput(861.00,407.27)(12.639,-5.000){2}{\rule{0.569pt}{0.350pt}}
\multiput(876.00,402.02)(1.831,-0.507){7}{\rule{1.208pt}{0.122pt}}
\multiput(876.00,402.27)(13.494,-5.000){2}{\rule{0.604pt}{0.350pt}}
\multiput(892.00,397.02)(1.376,-0.505){9}{\rule{0.962pt}{0.122pt}}
\multiput(892.00,397.27)(13.002,-6.000){2}{\rule{0.481pt}{0.350pt}}
\multiput(907.00,391.02)(1.376,-0.505){9}{\rule{0.962pt}{0.122pt}}
\multiput(907.00,391.27)(13.002,-6.000){2}{\rule{0.481pt}{0.350pt}}
\multiput(922.00,385.02)(1.232,-0.504){11}{\rule{0.887pt}{0.121pt}}
\multiput(922.00,385.27)(14.158,-7.000){2}{\rule{0.444pt}{0.350pt}}
\multiput(938.00,378.02)(0.993,-0.504){13}{\rule{0.744pt}{0.121pt}}
\multiput(938.00,378.27)(13.456,-8.000){2}{\rule{0.372pt}{0.350pt}}
\multiput(953.00,370.02)(0.873,-0.503){15}{\rule{0.671pt}{0.121pt}}
\multiput(953.00,370.27)(13.608,-9.000){2}{\rule{0.335pt}{0.350pt}}
\multiput(968.00,361.02)(0.779,-0.503){17}{\rule{0.612pt}{0.121pt}}
\multiput(968.00,361.27)(13.729,-10.000){2}{\rule{0.306pt}{0.350pt}}
\multiput(983.00,351.02)(0.629,-0.502){23}{\rule{0.518pt}{0.121pt}}
\multiput(983.00,351.27)(14.924,-13.000){2}{\rule{0.259pt}{0.350pt}}
\multiput(999.00,338.02)(0.507,-0.502){27}{\rule{0.438pt}{0.121pt}}
\multiput(999.00,338.27)(14.092,-15.000){2}{\rule{0.219pt}{0.350pt}}
\multiput(1014.48,321.60)(0.502,-0.715){27}{\rule{0.121pt}{0.577pt}}
\multiput(1013.27,322.80)(15.000,-19.801){2}{\rule{0.350pt}{0.289pt}}
\multiput(1029.48,299.73)(0.502,-1.028){27}{\rule{0.121pt}{0.787pt}}
\multiput(1028.27,301.37)(15.000,-28.366){2}{\rule{0.350pt}{0.394pt}}
\multiput(1044.48,266.73)(0.502,-2.098){29}{\rule{0.121pt}{1.509pt}}
\multiput(1043.27,269.87)(16.000,-61.867){2}{\rule{0.350pt}{0.755pt}}
\put(418.0,467.0){\rule[-0.175pt]{3.854pt}{0.350pt}}
\sbox{\plotpoint}{\rule[-0.300pt]{0.600pt}{0.600pt}}%
\put(266,559){\usebox{\plotpoint}}
\multiput(266.00,557.50)(0.575,-0.500){21}{\rule{0.842pt}{0.121pt}}
\multiput(266.00,557.75)(13.252,-13.000){2}{\rule{0.421pt}{0.600pt}}
\multiput(281.00,544.50)(1.726,-0.502){5}{\rule{1.950pt}{0.121pt}}
\multiput(281.00,544.75)(10.953,-5.000){2}{\rule{0.975pt}{0.600pt}}
\put(296,538.75){\rule{3.854pt}{0.600pt}}
\multiput(296.00,539.75)(8.000,-2.000){2}{\rule{1.927pt}{0.600pt}}
\put(312,537.25){\rule{3.614pt}{0.600pt}}
\multiput(312.00,537.75)(7.500,-1.000){2}{\rule{1.807pt}{0.600pt}}
\put(327,536.25){\rule{3.614pt}{0.600pt}}
\multiput(327.00,536.75)(7.500,-1.000){2}{\rule{1.807pt}{0.600pt}}
\put(342,535.25){\rule{3.614pt}{0.600pt}}
\multiput(342.00,535.75)(7.500,-1.000){2}{\rule{1.807pt}{0.600pt}}
\put(357,534.25){\rule{3.854pt}{0.600pt}}
\multiput(357.00,534.75)(8.000,-1.000){2}{\rule{1.927pt}{0.600pt}}
\put(388,533.25){\rule{3.614pt}{0.600pt}}
\multiput(388.00,533.75)(7.500,-1.000){2}{\rule{1.807pt}{0.600pt}}
\put(403,532.25){\rule{3.614pt}{0.600pt}}
\multiput(403.00,532.75)(7.500,-1.000){2}{\rule{1.807pt}{0.600pt}}
\put(418,531.25){\rule{3.854pt}{0.600pt}}
\multiput(418.00,531.75)(8.000,-1.000){2}{\rule{1.927pt}{0.600pt}}
\put(434,530.25){\rule{3.614pt}{0.600pt}}
\multiput(434.00,530.75)(7.500,-1.000){2}{\rule{1.807pt}{0.600pt}}
\put(449,529.25){\rule{3.614pt}{0.600pt}}
\multiput(449.00,529.75)(7.500,-1.000){2}{\rule{1.807pt}{0.600pt}}
\put(464,528.25){\rule{3.854pt}{0.600pt}}
\multiput(464.00,528.75)(8.000,-1.000){2}{\rule{1.927pt}{0.600pt}}
\put(480,527.25){\rule{3.614pt}{0.600pt}}
\multiput(480.00,527.75)(7.500,-1.000){2}{\rule{1.807pt}{0.600pt}}
\put(495,526.25){\rule{3.614pt}{0.600pt}}
\multiput(495.00,526.75)(7.500,-1.000){2}{\rule{1.807pt}{0.600pt}}
\put(510,525.25){\rule{3.614pt}{0.600pt}}
\multiput(510.00,525.75)(7.500,-1.000){2}{\rule{1.807pt}{0.600pt}}
\put(525,524.25){\rule{3.854pt}{0.600pt}}
\multiput(525.00,524.75)(8.000,-1.000){2}{\rule{1.927pt}{0.600pt}}
\put(541,523.25){\rule{3.614pt}{0.600pt}}
\multiput(541.00,523.75)(7.500,-1.000){2}{\rule{1.807pt}{0.600pt}}
\put(556,521.75){\rule{3.614pt}{0.600pt}}
\multiput(556.00,522.75)(7.500,-2.000){2}{\rule{1.807pt}{0.600pt}}
\put(571,520.25){\rule{3.614pt}{0.600pt}}
\multiput(571.00,520.75)(7.500,-1.000){2}{\rule{1.807pt}{0.600pt}}
\put(586,518.75){\rule{3.854pt}{0.600pt}}
\multiput(586.00,519.75)(8.000,-2.000){2}{\rule{1.927pt}{0.600pt}}
\put(602,516.75){\rule{3.614pt}{0.600pt}}
\multiput(602.00,517.75)(7.500,-2.000){2}{\rule{1.807pt}{0.600pt}}
\put(617,515.25){\rule{3.614pt}{0.600pt}}
\multiput(617.00,515.75)(7.500,-1.000){2}{\rule{1.807pt}{0.600pt}}
\put(632,513.75){\rule{3.614pt}{0.600pt}}
\multiput(632.00,514.75)(7.500,-2.000){2}{\rule{1.807pt}{0.600pt}}
\put(647,511.75){\rule{3.854pt}{0.600pt}}
\multiput(647.00,512.75)(8.000,-2.000){2}{\rule{1.927pt}{0.600pt}}
\put(663,509.75){\rule{3.614pt}{0.600pt}}
\multiput(663.00,510.75)(7.500,-2.000){2}{\rule{1.807pt}{0.600pt}}
\put(678,507.75){\rule{3.614pt}{0.600pt}}
\multiput(678.00,508.75)(7.500,-2.000){2}{\rule{1.807pt}{0.600pt}}
\put(693,505.25){\rule{3.350pt}{0.600pt}}
\multiput(693.00,506.75)(9.047,-3.000){2}{\rule{1.675pt}{0.600pt}}
\put(709,502.75){\rule{3.614pt}{0.600pt}}
\multiput(709.00,503.75)(7.500,-2.000){2}{\rule{1.807pt}{0.600pt}}
\put(724,500.25){\rule{3.150pt}{0.600pt}}
\multiput(724.00,501.75)(8.462,-3.000){2}{\rule{1.575pt}{0.600pt}}
\put(739,497.75){\rule{3.614pt}{0.600pt}}
\multiput(739.00,498.75)(7.500,-2.000){2}{\rule{1.807pt}{0.600pt}}
\put(754,495.25){\rule{3.350pt}{0.600pt}}
\multiput(754.00,496.75)(9.047,-3.000){2}{\rule{1.675pt}{0.600pt}}
\multiput(770.00,493.50)(2.519,-0.503){3}{\rule{2.400pt}{0.121pt}}
\multiput(770.00,493.75)(10.019,-4.000){2}{\rule{1.200pt}{0.600pt}}
\put(785,488.25){\rule{3.150pt}{0.600pt}}
\multiput(785.00,489.75)(8.462,-3.000){2}{\rule{1.575pt}{0.600pt}}
\put(800,485.25){\rule{3.150pt}{0.600pt}}
\multiput(800.00,486.75)(8.462,-3.000){2}{\rule{1.575pt}{0.600pt}}
\multiput(815.00,483.50)(2.707,-0.503){3}{\rule{2.550pt}{0.121pt}}
\multiput(815.00,483.75)(10.707,-4.000){2}{\rule{1.275pt}{0.600pt}}
\multiput(831.00,479.50)(1.726,-0.502){5}{\rule{1.950pt}{0.121pt}}
\multiput(831.00,479.75)(10.953,-5.000){2}{\rule{0.975pt}{0.600pt}}
\multiput(846.00,474.50)(2.519,-0.503){3}{\rule{2.400pt}{0.121pt}}
\multiput(846.00,474.75)(10.019,-4.000){2}{\rule{1.200pt}{0.600pt}}
\multiput(861.00,470.50)(1.726,-0.502){5}{\rule{1.950pt}{0.121pt}}
\multiput(861.00,470.75)(10.953,-5.000){2}{\rule{0.975pt}{0.600pt}}
\multiput(876.00,465.50)(1.852,-0.502){5}{\rule{2.070pt}{0.121pt}}
\multiput(876.00,465.75)(11.704,-5.000){2}{\rule{1.035pt}{0.600pt}}
\multiput(892.00,460.50)(1.358,-0.501){7}{\rule{1.650pt}{0.121pt}}
\multiput(892.00,460.75)(11.575,-6.000){2}{\rule{0.825pt}{0.600pt}}
\multiput(907.00,454.50)(1.358,-0.501){7}{\rule{1.650pt}{0.121pt}}
\multiput(907.00,454.75)(11.575,-6.000){2}{\rule{0.825pt}{0.600pt}}
\multiput(922.00,448.50)(1.211,-0.501){9}{\rule{1.521pt}{0.121pt}}
\multiput(922.00,448.75)(12.842,-7.000){2}{\rule{0.761pt}{0.600pt}}
\multiput(938.00,441.50)(0.852,-0.501){13}{\rule{1.150pt}{0.121pt}}
\multiput(938.00,441.75)(12.613,-9.000){2}{\rule{0.575pt}{0.600pt}}
\multiput(953.00,432.50)(0.852,-0.501){13}{\rule{1.150pt}{0.121pt}}
\multiput(953.00,432.75)(12.613,-9.000){2}{\rule{0.575pt}{0.600pt}}
\multiput(968.00,423.50)(0.760,-0.501){15}{\rule{1.050pt}{0.121pt}}
\multiput(968.00,423.75)(12.821,-10.000){2}{\rule{0.525pt}{0.600pt}}
\multiput(983.00,413.50)(0.616,-0.500){21}{\rule{0.888pt}{0.121pt}}
\multiput(983.00,413.75)(14.156,-13.000){2}{\rule{0.444pt}{0.600pt}}
\multiput(1000.00,398.72)(0.500,-0.530){25}{\rule{0.121pt}{0.790pt}}
\multiput(997.75,400.36)(15.000,-14.360){2}{\rule{0.600pt}{0.395pt}}
\multiput(1015.00,381.72)(0.500,-0.739){25}{\rule{0.121pt}{1.030pt}}
\multiput(1012.75,383.86)(15.000,-19.862){2}{\rule{0.600pt}{0.515pt}}
\multiput(1030.00,358.06)(0.500,-1.086){25}{\rule{0.121pt}{1.430pt}}
\multiput(1027.75,361.03)(15.000,-29.032){2}{\rule{0.600pt}{0.715pt}}
\multiput(1045.00,321.10)(0.500,-2.119){27}{\rule{0.121pt}{2.625pt}}
\multiput(1042.75,326.55)(16.000,-60.552){2}{\rule{0.600pt}{1.313pt}}
\put(373.0,535.0){\rule[-0.300pt]{3.613pt}{0.600pt}}
\end{picture}
\vspace{0.4cm}
\caption{Illustrative plot for $|V_{cb}|$ changing with 
$\tan\beta$ at the weak scale $M_Z$, obtained from the patterns of quark mass matrices
S1, S2, A1 and A2.}
\end{figure}

\vspace{1cm}

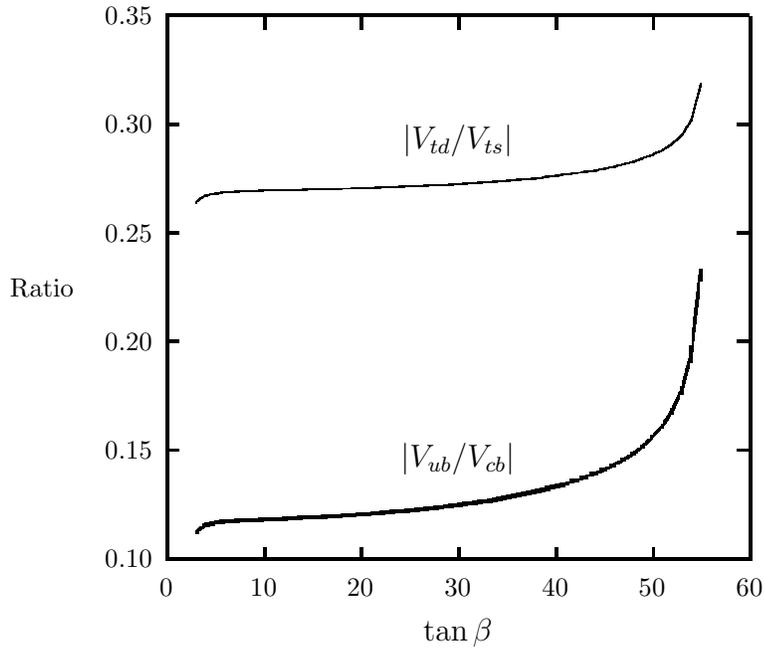
\begin{figure}
\setlength{\unitlength}{0.240900pt}
\ifx\plotpoint\undefined\newsavebox{\plotpoint}\fi
\sbox{\plotpoint}{\rule[-0.500pt]{1.000pt}{1.000pt}}%
\begin{picture}(1200,990)(-350,0)
\font\gnuplot=cmr10 at 10pt
\gnuplot
\sbox{\plotpoint}{\rule[-0.500pt]{1.000pt}{1.000pt}}%
\put(220.0,113.0){\rule[-0.500pt]{1.000pt}{205.729pt}}
\put(220.0,113.0){\rule[-0.500pt]{4.818pt}{1.000pt}}
\put(198,113){\makebox(0,0)[r]{0.10}}
\put(1116.0,113.0){\rule[-0.500pt]{4.818pt}{1.000pt}}
\put(220.0,284.0){\rule[-0.500pt]{4.818pt}{1.000pt}}
\put(198,284){\makebox(0,0)[r]{0.15}}
\put(1116.0,284.0){\rule[-0.500pt]{4.818pt}{1.000pt}}
\put(220.0,455.0){\rule[-0.500pt]{4.818pt}{1.000pt}}
\put(198,455){\makebox(0,0)[r]{0.20}}
\put(1116.0,455.0){\rule[-0.500pt]{4.818pt}{1.000pt}}
\put(220.0,625.0){\rule[-0.500pt]{4.818pt}{1.000pt}}
\put(198,625){\makebox(0,0)[r]{0.25}}
\put(1116.0,625.0){\rule[-0.500pt]{4.818pt}{1.000pt}}
\put(220.0,796.0){\rule[-0.500pt]{4.818pt}{1.000pt}}
\put(198,796){\makebox(0,0)[r]{0.30}}
\put(1116.0,796.0){\rule[-0.500pt]{4.818pt}{1.000pt}}
\put(220.0,967.0){\rule[-0.500pt]{4.818pt}{1.000pt}}
\put(198,967){\makebox(0,0)[r]{0.35}}
\put(1116.0,967.0){\rule[-0.500pt]{4.818pt}{1.000pt}}
\put(220.0,113.0){\rule[-0.500pt]{1.000pt}{4.818pt}}
\put(220,68){\makebox(0,0){0}}
\put(220.0,947.0){\rule[-0.500pt]{1.000pt}{4.818pt}}
\put(373.0,113.0){\rule[-0.500pt]{1.000pt}{4.818pt}}
\put(373,68){\makebox(0,0){10}}
\put(373.0,947.0){\rule[-0.500pt]{1.000pt}{4.818pt}}
\put(525.0,113.0){\rule[-0.500pt]{1.000pt}{4.818pt}}
\put(525,68){\makebox(0,0){20}}
\put(525.0,947.0){\rule[-0.500pt]{1.000pt}{4.818pt}}
\put(678.0,113.0){\rule[-0.500pt]{1.000pt}{4.818pt}}
\put(678,68){\makebox(0,0){30}}
\put(678.0,947.0){\rule[-0.500pt]{1.000pt}{4.818pt}}
\put(831.0,113.0){\rule[-0.500pt]{1.000pt}{4.818pt}}
\put(831,68){\makebox(0,0){40}}
\put(831.0,947.0){\rule[-0.500pt]{1.000pt}{4.818pt}}
\put(983.0,113.0){\rule[-0.500pt]{1.000pt}{4.818pt}}
\put(983,68){\makebox(0,0){50}}
\put(983.0,947.0){\rule[-0.500pt]{1.000pt}{4.818pt}}
\put(1136.0,113.0){\rule[-0.500pt]{1.000pt}{4.818pt}}
\put(1136,68){\makebox(0,0){60}}
\put(1136.0,947.0){\rule[-0.500pt]{1.000pt}{4.818pt}}
\put(220.0,113.0){\rule[-0.500pt]{220.664pt}{1.000pt}}
\put(1136.0,113.0){\rule[-0.500pt]{1.000pt}{205.729pt}}
\put(220.0,967.0){\rule[-0.500pt]{220.664pt}{1.000pt}}
\put(25,540){\makebox(0,0){Ratio}}
\put(678,-3){\makebox(0,0){$\tan\beta$}}
\put(678,770){\makebox(0,0){$|V_{td}/V_{ts}|$}}
\put(678,270){\makebox(0,0){$|V_{ub}/V_{cb}|$}}

\put(220.0,113.0){\rule[-0.500pt]{1.000pt}{205.729pt}}
\put(266,155){\usebox{\plotpoint}}
\multiput(266.00,156.83)(0.639,0.489){14}{\rule{1.614pt}{0.118pt}}
\multiput(266.00,152.92)(11.651,11.000){2}{\rule{0.807pt}{1.000pt}}
\put(281,165.92){\rule{3.614pt}{1.000pt}}
\multiput(281.00,163.92)(7.500,4.000){2}{\rule{1.807pt}{1.000pt}}
\put(296,168.92){\rule{3.854pt}{1.000pt}}
\multiput(296.00,167.92)(8.000,2.000){2}{\rule{1.927pt}{1.000pt}}
\put(312,170.42){\rule{3.614pt}{1.000pt}}
\multiput(312.00,169.92)(7.500,1.000){2}{\rule{1.807pt}{1.000pt}}
\put(327,171.42){\rule{3.614pt}{1.000pt}}
\multiput(327.00,170.92)(7.500,1.000){2}{\rule{1.807pt}{1.000pt}}
\put(357,172.42){\rule{3.854pt}{1.000pt}}
\multiput(357.00,171.92)(8.000,1.000){2}{\rule{1.927pt}{1.000pt}}
\put(373,173.42){\rule{3.614pt}{1.000pt}}
\multiput(373.00,172.92)(7.500,1.000){2}{\rule{1.807pt}{1.000pt}}
\put(342.0,174.0){\rule[-0.500pt]{3.613pt}{1.000pt}}
\put(403,174.42){\rule{3.614pt}{1.000pt}}
\multiput(403.00,173.92)(7.500,1.000){2}{\rule{1.807pt}{1.000pt}}
\put(418,175.42){\rule{3.854pt}{1.000pt}}
\multiput(418.00,174.92)(8.000,1.000){2}{\rule{1.927pt}{1.000pt}}
\put(434,176.42){\rule{3.614pt}{1.000pt}}
\multiput(434.00,175.92)(7.500,1.000){2}{\rule{1.807pt}{1.000pt}}
\put(388.0,176.0){\rule[-0.500pt]{3.613pt}{1.000pt}}
\put(464,177.42){\rule{3.854pt}{1.000pt}}
\multiput(464.00,176.92)(8.000,1.000){2}{\rule{1.927pt}{1.000pt}}
\put(480,178.42){\rule{3.614pt}{1.000pt}}
\multiput(480.00,177.92)(7.500,1.000){2}{\rule{1.807pt}{1.000pt}}
\put(495,179.42){\rule{3.614pt}{1.000pt}}
\multiput(495.00,178.92)(7.500,1.000){2}{\rule{1.807pt}{1.000pt}}
\put(510,180.42){\rule{3.614pt}{1.000pt}}
\multiput(510.00,179.92)(7.500,1.000){2}{\rule{1.807pt}{1.000pt}}
\put(525,181.42){\rule{3.854pt}{1.000pt}}
\multiput(525.00,180.92)(8.000,1.000){2}{\rule{1.927pt}{1.000pt}}
\put(541,182.42){\rule{3.614pt}{1.000pt}}
\multiput(541.00,181.92)(7.500,1.000){2}{\rule{1.807pt}{1.000pt}}
\put(556,183.92){\rule{3.614pt}{1.000pt}}
\multiput(556.00,182.92)(7.500,2.000){2}{\rule{1.807pt}{1.000pt}}
\put(571,185.42){\rule{3.614pt}{1.000pt}}
\multiput(571.00,184.92)(7.500,1.000){2}{\rule{1.807pt}{1.000pt}}
\put(586,186.42){\rule{3.854pt}{1.000pt}}
\multiput(586.00,185.92)(8.000,1.000){2}{\rule{1.927pt}{1.000pt}}
\put(602,187.92){\rule{3.614pt}{1.000pt}}
\multiput(602.00,186.92)(7.500,2.000){2}{\rule{1.807pt}{1.000pt}}
\put(617,189.42){\rule{3.614pt}{1.000pt}}
\multiput(617.00,188.92)(7.500,1.000){2}{\rule{1.807pt}{1.000pt}}
\put(632,190.92){\rule{3.614pt}{1.000pt}}
\multiput(632.00,189.92)(7.500,2.000){2}{\rule{1.807pt}{1.000pt}}
\put(647,192.92){\rule{3.854pt}{1.000pt}}
\multiput(647.00,191.92)(8.000,2.000){2}{\rule{1.927pt}{1.000pt}}
\put(663,194.92){\rule{3.614pt}{1.000pt}}
\multiput(663.00,193.92)(7.500,2.000){2}{\rule{1.807pt}{1.000pt}}
\put(678,196.92){\rule{3.614pt}{1.000pt}}
\multiput(678.00,195.92)(7.500,2.000){2}{\rule{1.807pt}{1.000pt}}
\put(693,198.92){\rule{3.854pt}{1.000pt}}
\multiput(693.00,197.92)(8.000,2.000){2}{\rule{1.927pt}{1.000pt}}
\put(709,200.92){\rule{3.614pt}{1.000pt}}
\multiput(709.00,199.92)(7.500,2.000){2}{\rule{1.807pt}{1.000pt}}
\put(724,203.42){\rule{3.614pt}{1.000pt}}
\multiput(724.00,201.92)(7.500,3.000){2}{\rule{1.807pt}{1.000pt}}
\put(739,206.42){\rule{3.614pt}{1.000pt}}
\multiput(739.00,204.92)(7.500,3.000){2}{\rule{1.807pt}{1.000pt}}
\put(754,209.42){\rule{3.854pt}{1.000pt}}
\multiput(754.00,207.92)(8.000,3.000){2}{\rule{1.927pt}{1.000pt}}
\put(770,212.42){\rule{3.614pt}{1.000pt}}
\multiput(770.00,210.92)(7.500,3.000){2}{\rule{1.807pt}{1.000pt}}
\put(785,215.42){\rule{3.614pt}{1.000pt}}
\multiput(785.00,213.92)(7.500,3.000){2}{\rule{1.807pt}{1.000pt}}
\put(800,218.92){\rule{3.614pt}{1.000pt}}
\multiput(800.00,216.92)(7.500,4.000){2}{\rule{1.807pt}{1.000pt}}
\put(815,222.92){\rule{3.854pt}{1.000pt}}
\multiput(815.00,220.92)(8.000,4.000){2}{\rule{1.927pt}{1.000pt}}
\put(831,226.92){\rule{3.614pt}{1.000pt}}
\multiput(831.00,224.92)(7.500,4.000){2}{\rule{1.807pt}{1.000pt}}
\multiput(846.00,232.86)(1.509,0.424){2}{\rule{3.250pt}{0.102pt}}
\multiput(846.00,228.92)(8.254,5.000){2}{\rule{1.625pt}{1.000pt}}
\multiput(861.00,237.86)(1.509,0.424){2}{\rule{3.250pt}{0.102pt}}
\multiput(861.00,233.92)(8.254,5.000){2}{\rule{1.625pt}{1.000pt}}
\multiput(876.00,242.84)(1.298,0.462){4}{\rule{2.917pt}{0.111pt}}
\multiput(876.00,238.92)(9.946,6.000){2}{\rule{1.458pt}{1.000pt}}
\multiput(892.00,248.84)(1.195,0.462){4}{\rule{2.750pt}{0.111pt}}
\multiput(892.00,244.92)(9.292,6.000){2}{\rule{1.375pt}{1.000pt}}
\multiput(907.00,254.84)(1.013,0.475){6}{\rule{2.393pt}{0.114pt}}
\multiput(907.00,250.92)(10.034,7.000){2}{\rule{1.196pt}{1.000pt}}
\multiput(922.00,261.83)(0.842,0.485){10}{\rule{2.028pt}{0.117pt}}
\multiput(922.00,257.92)(11.791,9.000){2}{\rule{1.014pt}{1.000pt}}
\multiput(938.00,270.83)(0.783,0.485){10}{\rule{1.917pt}{0.117pt}}
\multiput(938.00,266.92)(11.022,9.000){2}{\rule{0.958pt}{1.000pt}}
\multiput(953.00,279.83)(0.585,0.491){16}{\rule{1.500pt}{0.118pt}}
\multiput(953.00,275.92)(11.887,12.000){2}{\rule{0.750pt}{1.000pt}}
\multiput(968.00,291.83)(0.501,0.492){20}{\rule{1.321pt}{0.119pt}}
\multiput(968.00,287.92)(12.257,14.000){2}{\rule{0.661pt}{1.000pt}}
\multiput(984.83,304.00)(0.494,0.502){24}{\rule{0.119pt}{1.312pt}}
\multiput(980.92,304.00)(16.000,14.276){2}{\rule{1.000pt}{0.656pt}}
\multiput(1000.83,321.00)(0.493,0.743){22}{\rule{0.119pt}{1.783pt}}
\multiput(996.92,321.00)(15.000,19.299){2}{\rule{1.000pt}{0.892pt}}
\multiput(1015.83,344.00)(0.493,1.053){22}{\rule{0.119pt}{2.383pt}}
\multiput(1011.92,344.00)(15.000,27.053){2}{\rule{1.000pt}{1.192pt}}
\multiput(1030.83,376.00)(0.493,1.810){22}{\rule{0.119pt}{3.850pt}}
\multiput(1026.92,376.00)(15.000,46.009){2}{\rule{1.000pt}{1.925pt}}
\multiput(1045.83,430.00)(0.494,4.398){24}{\rule{0.119pt}{8.875pt}}
\multiput(1041.92,430.00)(16.000,119.579){2}{\rule{1.000pt}{4.438pt}}
\put(449.0,179.0){\rule[-0.500pt]{3.613pt}{1.000pt}}
\sbox{\plotpoint}{\rule[-0.175pt]{0.350pt}{0.350pt}}%
\put(266,673){\usebox{\plotpoint}}
\multiput(266.00,673.48)(0.703,0.502){19}{\rule{0.565pt}{0.121pt}}
\multiput(266.00,672.27)(13.828,11.000){2}{\rule{0.282pt}{0.350pt}}
\multiput(281.00,684.47)(3.686,0.516){3}{\rule{1.837pt}{0.124pt}}
\multiput(281.00,683.27)(11.186,3.000){2}{\rule{0.919pt}{0.350pt}}
\put(296,687.27){\rule{2.888pt}{0.350pt}}
\multiput(296.00,686.27)(10.007,2.000){2}{\rule{1.444pt}{0.350pt}}
\put(312,688.77){\rule{3.614pt}{0.350pt}}
\multiput(312.00,688.27)(7.500,1.000){2}{\rule{1.807pt}{0.350pt}}
\put(327,689.77){\rule{3.614pt}{0.350pt}}
\multiput(327.00,689.27)(7.500,1.000){2}{\rule{1.807pt}{0.350pt}}
\put(357,690.77){\rule{3.854pt}{0.350pt}}
\multiput(357.00,690.27)(8.000,1.000){2}{\rule{1.927pt}{0.350pt}}
\put(342.0,691.0){\rule[-0.175pt]{3.613pt}{0.350pt}}
\put(403,691.77){\rule{3.614pt}{0.350pt}}
\multiput(403.00,691.27)(7.500,1.000){2}{\rule{1.807pt}{0.350pt}}
\put(373.0,692.0){\rule[-0.175pt]{7.227pt}{0.350pt}}
\put(434,692.77){\rule{3.614pt}{0.350pt}}
\multiput(434.00,692.27)(7.500,1.000){2}{\rule{1.807pt}{0.350pt}}
\put(418.0,693.0){\rule[-0.175pt]{3.854pt}{0.350pt}}
\put(480,693.77){\rule{3.614pt}{0.350pt}}
\multiput(480.00,693.27)(7.500,1.000){2}{\rule{1.807pt}{0.350pt}}
\put(449.0,694.0){\rule[-0.175pt]{7.468pt}{0.350pt}}
\put(510,694.77){\rule{3.614pt}{0.350pt}}
\multiput(510.00,694.27)(7.500,1.000){2}{\rule{1.807pt}{0.350pt}}
\put(495.0,695.0){\rule[-0.175pt]{3.613pt}{0.350pt}}
\put(541,695.77){\rule{3.614pt}{0.350pt}}
\multiput(541.00,695.27)(7.500,1.000){2}{\rule{1.807pt}{0.350pt}}
\put(525.0,696.0){\rule[-0.175pt]{3.854pt}{0.350pt}}
\put(571,696.77){\rule{3.614pt}{0.350pt}}
\multiput(571.00,696.27)(7.500,1.000){2}{\rule{1.807pt}{0.350pt}}
\put(586,697.77){\rule{3.854pt}{0.350pt}}
\multiput(586.00,697.27)(8.000,1.000){2}{\rule{1.927pt}{0.350pt}}
\put(556.0,697.0){\rule[-0.175pt]{3.613pt}{0.350pt}}
\put(617,698.77){\rule{3.614pt}{0.350pt}}
\multiput(617.00,698.27)(7.500,1.000){2}{\rule{1.807pt}{0.350pt}}
\put(632,699.77){\rule{3.614pt}{0.350pt}}
\multiput(632.00,699.27)(7.500,1.000){2}{\rule{1.807pt}{0.350pt}}
\put(602.0,699.0){\rule[-0.175pt]{3.613pt}{0.350pt}}
\put(663,700.77){\rule{3.614pt}{0.350pt}}
\multiput(663.00,700.27)(7.500,1.000){2}{\rule{1.807pt}{0.350pt}}
\put(678,701.77){\rule{3.614pt}{0.350pt}}
\multiput(678.00,701.27)(7.500,1.000){2}{\rule{1.807pt}{0.350pt}}
\put(693,702.77){\rule{3.854pt}{0.350pt}}
\multiput(693.00,702.27)(8.000,1.000){2}{\rule{1.927pt}{0.350pt}}
\put(709,703.77){\rule{3.614pt}{0.350pt}}
\multiput(709.00,703.27)(7.500,1.000){2}{\rule{1.807pt}{0.350pt}}
\put(724,704.77){\rule{3.614pt}{0.350pt}}
\multiput(724.00,704.27)(7.500,1.000){2}{\rule{1.807pt}{0.350pt}}
\put(739,705.77){\rule{3.614pt}{0.350pt}}
\multiput(739.00,705.27)(7.500,1.000){2}{\rule{1.807pt}{0.350pt}}
\put(754,707.27){\rule{2.888pt}{0.350pt}}
\multiput(754.00,706.27)(10.007,2.000){2}{\rule{1.444pt}{0.350pt}}
\put(770,708.77){\rule{3.614pt}{0.350pt}}
\multiput(770.00,708.27)(7.500,1.000){2}{\rule{1.807pt}{0.350pt}}
\put(785,709.77){\rule{3.614pt}{0.350pt}}
\multiput(785.00,709.27)(7.500,1.000){2}{\rule{1.807pt}{0.350pt}}
\put(800,711.27){\rule{2.712pt}{0.350pt}}
\multiput(800.00,710.27)(9.370,2.000){2}{\rule{1.356pt}{0.350pt}}
\put(815,713.27){\rule{2.888pt}{0.350pt}}
\multiput(815.00,712.27)(10.007,2.000){2}{\rule{1.444pt}{0.350pt}}
\put(831,715.27){\rule{2.712pt}{0.350pt}}
\multiput(831.00,714.27)(9.370,2.000){2}{\rule{1.356pt}{0.350pt}}
\put(846,717.27){\rule{2.712pt}{0.350pt}}
\multiput(846.00,716.27)(9.370,2.000){2}{\rule{1.356pt}{0.350pt}}
\put(861,719.27){\rule{2.712pt}{0.350pt}}
\multiput(861.00,718.27)(9.370,2.000){2}{\rule{1.356pt}{0.350pt}}
\put(876,721.27){\rule{2.888pt}{0.350pt}}
\multiput(876.00,720.27)(10.007,2.000){2}{\rule{1.444pt}{0.350pt}}
\multiput(892.00,723.47)(3.686,0.516){3}{\rule{1.837pt}{0.124pt}}
\multiput(892.00,722.27)(11.186,3.000){2}{\rule{0.919pt}{0.350pt}}
\multiput(907.00,726.47)(3.686,0.516){3}{\rule{1.837pt}{0.124pt}}
\multiput(907.00,725.27)(11.186,3.000){2}{\rule{0.919pt}{0.350pt}}
\multiput(922.00,729.47)(2.456,0.509){5}{\rule{1.487pt}{0.123pt}}
\multiput(922.00,728.27)(12.913,4.000){2}{\rule{0.744pt}{0.350pt}}
\multiput(938.00,733.47)(2.297,0.509){5}{\rule{1.400pt}{0.123pt}}
\multiput(938.00,732.27)(12.094,4.000){2}{\rule{0.700pt}{0.350pt}}
\multiput(953.00,737.47)(1.713,0.507){7}{\rule{1.137pt}{0.122pt}}
\multiput(953.00,736.27)(12.639,5.000){2}{\rule{0.569pt}{0.350pt}}
\multiput(968.00,742.47)(1.376,0.505){9}{\rule{0.962pt}{0.122pt}}
\multiput(968.00,741.27)(13.002,6.000){2}{\rule{0.481pt}{0.350pt}}
\multiput(983.00,748.47)(1.232,0.504){11}{\rule{0.887pt}{0.121pt}}
\multiput(983.00,747.27)(14.158,7.000){2}{\rule{0.444pt}{0.350pt}}
\multiput(999.00,755.48)(0.779,0.503){17}{\rule{0.612pt}{0.121pt}}
\multiput(999.00,754.27)(13.729,10.000){2}{\rule{0.306pt}{0.350pt}}
\multiput(1014.00,765.48)(0.589,0.502){23}{\rule{0.491pt}{0.121pt}}
\multiput(1014.00,764.27)(13.980,13.000){2}{\rule{0.246pt}{0.350pt}}
\multiput(1029.48,778.00)(0.502,0.785){27}{\rule{0.121pt}{0.624pt}}
\multiput(1028.27,778.00)(15.000,21.705){2}{\rule{0.350pt}{0.312pt}}
\multiput(1044.48,801.00)(0.502,1.903){29}{\rule{0.121pt}{1.378pt}}
\multiput(1043.27,801.00)(16.000,56.140){2}{\rule{0.350pt}{0.689pt}}
\put(647.0,701.0){\rule[-0.175pt]{3.854pt}{0.350pt}}
\end{picture}
\vspace{0.4cm}
\caption{Illustrative plot for the ratios $|V_{ub}/V_{cb}|$ and
$|V_{td}/V_{ts}|$ changing with $\tan\beta$ at the weak scale
$M_Z$, obtained from the quark mass matrix pattern A2 (here $\phi
=\pi/2$ has been taken).}
\end{figure}


\newpage

\begin{thebibliography}{99}

\bibitem{qmass}
C.D. Froggatt and H.B. Nielsen, Nucl. Phys. {\bf B 147}, 277 (1979); 
S. Dimopoulos, Phys. Lett. {\bf B 129}, 417 (1983).

\bibitem{IR}
L.E. Ib\'a\~nez and G.G. Ross, Phys. Lett. {\bf B 332}, 100 (1994).

\bibitem{qmass2}
M. Leurer, Y. Nir and N. Seiberg, Nucl. Phys. {\bf B 398}, 319 (1993); 
Nucl. Phys. {\bf B 420}, 468 (1994); 
Y. Nir and N. Seiberg, Phys. Lett. {\bf B 309}, 337 (1993);
V.~Jain and R.~Shrock, Phys. Lett. {\bf B 352}, 83 (1995);
P.~Bin\'etruy and P.~Ramond, Phys. Lett. {\bf B 350}, 49 (1995);
E.~Dudas, S.~Pokorski and C.A.~Savoy, Phys. Lett. {\bf B 356}, 45 (1995).

\bibitem{Dterm}
M.~Drees, Phys. Lett. {\bf B 181}, 279 (1986);
J.S.~Hagelin and S.~Kelley, Nucl. Phys. {\bf B 342}, 95 (1990);
A.E.~Faraggi, J.S.~Hagelin, S.~Kelley and D.V.~Nanopoulos, 
Phys. Rev. {\bf D 45}, 3272 (1992);
Y.~Kawamura, Phys.~Rev.~{\bf D 53}, 3779 (1996);
Y.~Kawamura and T.~Kobayashi, Phys. Lett. {\bf B 375}, 141 (1996); 
Report No. INS-Rep-1153 (hep-ph/9608233); 
Y.~Kawamura, T.~Kobayashi and T.~Komatsu, 
Report No. INS-Rep-1161 (hep-ph/9609462), to be published 
in Phys. Lett. {\bf B}.

\bibitem{fcnc}
J. Ellis and D.V. Nanopoulos,  Phys. Lett. {\bf B 110}, 44 (1982);
R. Barbieri and R. Gatto, Phys. Lett. {\bf B 110}, 211 (1982);
T. Inami and C.S. Lim, Nucl. Phys.  {\bf B 207}, 533 (1982);
J.~Hagelin, S.~Kelly and T.~Tanaka, Nucl.~Phys. {\bf B 415}, 293 (1994).

\bibitem{stringm}
J.L.~Lopez and D.V.~Nanopoulos, Nucl. Phys. {\bf B 338}, 73 (1990);
A.E. Faraggi and E. Halyo, Nucl. Phys. {\bf B 416}, 63 (1994).

\bibitem{stringmcy}
N. Haba, C. Hattori, M. Matsuda, and T. Matsuoka,
Prog. Theor. Phys. {\bf 96}, 1249 (1996).

\bibitem{stringm3}
K.S.~Babu and R.N.~Mohapatra, Phys. Rev. Lett. {\bf 74}, 2418 (1995).

\bibitem{stringm21}
T.~Kobayashi, Phys. Lett. {\bf B 358}, 253 (1995).

\bibitem{stringm22}
T.~Kobayashi and Z.Z.~Xing, Mod. Phys. Lett. {\bf A 12}, 561 (1997).

\bibitem{nr}
M.~Cvetic, Phys. Rev. Lett. {\bf 59}, 1795 (1987);
A.~Font, L.E.~Ib\'a\~nez, H.P.~Nilles and F.~Quevedo, 
Phys. Lett. {\bf B 213}, 274 (1988).

\bibitem{NR} 
T. Kobayashi, Phys. Lett. {\bf B 354}, 264 (1995).

\bibitem{RRR}
R. Ramond, R.G. Roberts and G.G. Ross, Nucl. Phys. {\bf B 406}, 19 (1993). 

\bibitem{Orbi}
L.~Dixon, J.~Harvey, C.~Vafa and E.~Witten,
Nucl.~Phys. {\bf B 261}, 678 (1985); Nucl.~Phys. {\bf B 274}, 285 (1986);
L.E.~Ib\'a\~nez, J.~Mas, H.P.~Nilles and F.~Quevedo,
Nucl.~Phys. {\bf B 301}, 157 (1988);
Y.~Katsuki, Y.~Kawamura, T.~Kobayashi, N.~Ohtsubo, Y.~Ono and 
K.~Tanioka, Nucl.~Phys. {\bf B 341}, 611 (1990).

\bibitem{KO1}
T.~Kobayashi and N.~Ohtsubo, Phys.~Lett. {\bf B 245}, 441 (1990).

\bibitem{KO2}
T.~Kobayashi and N.~Ohtsubo, Int.~J.~Mod.~Phys. {\bf A 9}, 87 (1994).

\bibitem{FMS}
D.~Friedan, E.~Martinec and S.~Shenker, Nucl.~Phys. {\bf B 271}, 93 (1986).

\bibitem{Yukawa}
S.~Hamidi and C.~Vafa, Nucl.~Phys. {\bf B 279}, 465 (1987); 
L.~Dixon, D.~Friedan, E.~Martinec and S.~Shenker, Nucl.~Phys.
{\bf B 282}, 13 (1987).

\bibitem{Z6}
Y.~Katsuki, Y.~Kawamura, T.~Kobayashi, N.~Ohtsubo, Y.~Ono and 
K.~Tanioka, Phys. Lett. {\bf B 218}, 169 (1989);
Y.~Kawamura and T.~Kobayashi, Nucl.~Phys. {\bf B 481}, 539 (1996).

\bibitem{Fritzsch78} H. Fritzsch, Phys. Lett. {\bf B 73}, 317 (1978);
Nucl. Phys. {\bf B 155}, 189 (1979).

\bibitem{anti}
J. Erler, D. Jungnickel and J. Lauer, Phys. Rev. {\bf D 45}, 3651 (1992); 
D. Jungnickel, J. Lauer, M. Spali\'nski and S. Stieberger, 
Mod. Phys. Lett. {\bf A 7}, 3059 (1992); 
J. Erler, D. Jungnickel, M. Spali\'nski and S. Stieberger, 
Nucl. Phys. {\bf B 397}, 379 (1993).

\bibitem{CP}
C.S. Lim, Phys. Lett. {\bf B 256}, 233 (1991);
M. Dine R.G. Leigh and D.A. MacIntire, 
Phys. Rev. Lett. {\bf 69}, 2030 (1992);
K. Choi, D.B. Kaplan and A.E. Nelson, Nucl. Phys. {\bf B 391}, 515 (1993);
T. Kobayashi and C.S. Lim, Phys. Lett. {\bf B 343}, 122 (1995).

\bibitem{XingJPG} See, e.g., Z.Z. Xing, J. Phys. {\bf G 23}, 717 (1997).

\bibitem{PDG96} Particle Data Group, R.M. Barnett {\it et al.}, Phys. Rev.
{\bf D 54}, 1 (1996).

\bibitem{XingNPB} Z.Z. Xing, Nucl. Phys. {\bf B} (Proc. Suppl.) {\bf 50},
24 (1996); Nuovo Cimento {\bf A 109}, 115 (1996).

\bibitem{Babu93} K.S. Babu and Q. Shafi, Phys. Rev. {\bf D 47}, 5004 (1993).

\bibitem{Leutwyler96} H. Leutwyler, Phys. Lett. {\bf B 378}, 313 (1996);
and private communications.

\bibitem{FritzschXing95} H. Fritzsch and Z.Z. Xing, Phys. Lett. {\bf B 353},
114 (1995); Phys. Lett. {\bf B 413}, 396 (1997); hep-ph/9707215
and hep-ph/9708366.

\bibitem{Hall93} See, e.g., L.J. Hall and A. Rasin, Phys. Lett. {\bf B 315},
164 (1993); D. Du and Z.Z. Xing, Phys. Rev. {\bf D 48}, 2349 (1993);
S.S. Xue, Phys. Lett. {\bf B 398}, 177 (1997);
K. Harayama, N. Okamura, A.I. Sanda, and Z.Z. Xing, 
Prog. Theor. Phys. {\bf 97}, 781 (1997).

\bibitem{Gasser82} J. Gasser and H. Leutwyler, Phys. Rep. {\bf C 87}, 77 (1982).

\bibitem{Ali96} A. Ali and D. London, Report No. DESY 96-140 (talk presented 
at the QCD Euroconference 96, Montpellier, July, 1996).

\bibitem{Xing96} Z.Z. Xing, J. Phys. {\bf G 23}, 1563 (1997); hep-ph/9609204.

\end{thebibliography}
\end{document}